\documentclass[aps,pre,amsmath,amssymb,lengthcheck,showpacs,superscriptaddress]{revtex4-1}
\usepackage{graphicx}
\usepackage{color}
\usepackage{amsfonts}
\usepackage{physics}
\usepackage{braket}
\usepackage[utf8]{inputenc}

\begin{document}

\title{Novel elastic instability of amorphous solids in finite spatial dimensions}

\author{Masanari Shimada}
\email{masanari-shimada444@g.ecc.u-tokyo.ac.jp}
\affiliation{Graduate School of Arts and Sciences, The University of Tokyo, Tokyo 153-8902, Japan}
\author{Hideyuki Mizuno}
\affiliation{Graduate School of Arts and Sciences, The University of Tokyo, Tokyo 153-8902, Japan}
\author{Atsushi Ikeda}
\affiliation{Graduate School of Arts and Sciences, The University of Tokyo, Tokyo 153-8902, Japan}
\affiliation{Research Center for Complex Systems Biology, Universal Biology Institute, The University of Tokyo, Tokyo 153-8902, Japan}

\date{\today}

\begin{abstract}
Recently, progress has been made in the understanding of anomalous vibrational excitations in amorphous solids.
In the lowest-frequency region, the vibrational spectrum follows a non-Debye quartic law, which persists up to zero frequency without any frequency gap.
This gapless vibrational density of states (vDOS) suggests that glasses are on the verge of instability.
This feature of marginal stability is now highlighted as a key concept in the theories of glasses.
In particular, the elasticity theory based on marginal stability predicts the gapless vDOS. However, this theory yields a quadratic law and \textit{not} the quartic law.
To address this inconsistency, we presented a new type of instability, which is different from the conventional one, and proposed that amorphous solids are marginally stable considering the new instability in the preceding study~[M. Shimada, H. Mizuno, and A. Ikeda, Soft Matter, {\bf 16}, 7279, 2020].
In this study, we further extend and detail the results for these instabilities.
By analyzing various examples of disorder, we demonstrate that real glasses in finite spatial dimensions can be marginally stable by the proposed novel instability.
\end{abstract}

\maketitle

\section{Introduction}

Lattice vibrations of crystals, called phonons, can be fully described in terms of the spatial periodicity and defects~\cite{Kittel1996Introduction}.
The transportation of phonons controls the thermal properties, whereas structural defects cause mechanical failure.
In contrast, amorphous solids have no periodicity in their structures, and it is impossible to define defects unambiguously.
In amorphous solids, two species of anomalous vibrational modes have been observed in addition to phonons, even in the low-frequency region, where we can safely apply the Debye theory to crystals~\cite{Buchenau1984Neutron,Laird1991Localized,Schober1991Localized,Leonforte2005Continuum,Monaco2009Anomalous}.
The first species are heterogeneous and spatially extended vibrations.
They manifest as a peak at approximately $1$ THz in the vibrational density of states (vDOS) $g(\omega)$ divided by the squared frequency $\omega^2$, referred to as the boson peak (BP)~\cite{Buchenau1984Neutron}.
The second species are strongly anharmonic~\cite{Taraskin1999Anharmonicity,Xu2010Anharmonic} and are spatially localized vibrations referred to as quasilocalized vibrations (QLVs). 
These vibrational modes control the low-temperature thermal properties of glasses~\cite{Zeller1971Thermal,Anderson1972Anomalous,Phillips1972Tunneling,Phillips1981Amorphous,Karpov1983Theory,Buchenau1991Anharmonic,Buchenau1992Interaction}.
Because the frequency of the QLVs is significantly lower than the BP frequency, they affect mechanical failure~\cite{Maloney2006,Tanguy2010,Manning2011} under a load, as well as the structural relaxation of supercooled liquids near the glass transition temperature~\cite{Oligschleger1999,Widmer-Cooper2009}.
Therefore, the anomalous vibrations of glasses have attracted a vast interest in the past decades.

Recently, many numerical studies have reported the quantitative properties of these anomalous vibrations.
Simulations using weakly coordinated jammed packings near the jamming transition~\cite{Charbonneau2016Universal} have established that the vDOS obeys a power-law dependence, $g(\omega)\sim\omega^2$, at approximately the BP frequency. 
This scaling is independent of the spatial dimension $d$, which is distinct from the Debye law, $g_{\mathrm{Debye}}(\omega)\sim\omega^{d-1}$ and is referred to as the non-Debye scaling law.
Numerical studies further suggested that this scaling can also persist in Lennard–Jones glasses~\cite{Shimada2018Anomalous} far from the jamming transition in large spatial dimensions~\cite{Shimada_2020large}.

The non-Debye scaling law, however, does not extend down to zero frequency; instead, the coexistence of phonons and QLVs emerges below the BP frequency~\cite{Lerner2016Statistics,Mizuno2017Continuum,Wang2019Lowfrequency}.
The QLVs consist of a core and a far field that decays algebraically in space if they are not hybridized with phonons~\cite{Lerner2016Statistics,Gartner2016Nonlinear}.
This decay is sufficiently rapid for their participation ratio to scale as $1/N$ similar to truly localized vibrations, where $N$ is the number of particles.
A numerical study established that the motions of particles in the cores are energetically unstable, which are stabilized by the far-field components~\cite{Shimada2018Spatial}.
Moreover, the QLVs are similar to the response of a local dipolar force~\cite{Lerner2014Breakdown,Yan2016On,Shimada2018Spatial,Lerner18}, and their characteristic frequency increases rapidly as the glass transition is approached, along with a measure of elastic stiffness based on the local response~\cite{Lerner18}.
The vDOS of the QLVs follows another power law, $g_{\mathrm{QLV}}(\omega)\sim\omega^\beta$, where usually $\beta=4$~\cite{Lerner2016Statistics, Mizuno2017Continuum, Lerner2017Effect, Lerner2020Finite-size}, but $\beta\simeq3$ has been observed in small systems~\cite{Lerner2017Effect, Lerner2020Finite-size}.
Note that the vDOS power law persists at zero frequency.
That is, the vDOS of the QLVs is {\it gapless}, whereas the non-Debye scaling law is {\it gapped} in finite dimensions.

The gapless nature of the QLVs is significantly reminiscent of the marginal stability of glasses~\cite{Muller2015Marginal}. Amorphous systems are susceptible to infinitesimal perturbations such as shear deformation and thermal agitation.
A numerical study regarding the yielding transition found that glasses yield under infinitesimally small strains in the thermodynamic limit~\cite{Karmakar2010Statistical}.
Similarly, intermittent rearrangements are induced in glasses by infinitesimal thermal energy~\cite{Mizuno2020anharmonic}.

This marginality of glasses can be rationalized as follows~\cite{Wyart2005Effects, Muller2015Marginal}.
Generally, the configurational space explored by a system can be divided into the following three categories in terms of the stability of elementary excitations: absolutely stable, unstable, and marginally stable configurations~\cite{Wyart2005Effects, Muller2015Marginal}.
When we perturb an absolutely stable configuration, it is forced to return to the initial point by restoring forces, whereas an unstable configuration moves away from the initial point~\cite{Wyart2005Effects, Muller2015Marginal}.
The marginally stable phase exists between the two phases~\cite{Wyart2005Effects, Muller2015Marginal}.
Specifically, for particulate systems, normal liquids are unstable and frequently undergo structural relaxations, whereas crystals are stable.
Unlike crystals, glasses are predicted to be marginally stable considering their dynamics.
When we prepare glasses, normal liquids are usually quenched.
 First, the dynamics are driven by unstable excitations, and structural relaxations occur.
When the system approaches the stable phase, the number of unstable excitations decreases, and when it attains marginal stability, the dynamics freezes~\cite{Wyart2005Effects, Muller2015Marginal}.
Note that the marginal stability requires low-energy gapless excitations~\cite{Muller2015Marginal}, which may be identified as QLVs in the case of glasses.
\footnote{
In general, the concept of marginal stability requires understanding other phenomena, including crackling in a finite range of external fields~\cite{Muller2015Marginal}.
We do not consider these related topics in this study.
}.

Considering marginal stability, several attempts have been made to present the low-frequency vibrations of glasses~\cite{Schirmacher2006Thermal,Schirmacher2007Acoustic,Wyart2010Scaling, Degiuli2014Effects,Franz2015Universal,Ikeda2018Note,Ikeda2018Universal}.
Among them, we focus on the elasticity theory with a quenched disorder ~\cite{Schirmacher2006Thermal,Schirmacher2007Acoustic,Wyart2010Scaling, Degiuli2014Effects}.
The theory analyzes the elasticity model, either a coarse-grained continuum~\cite{Schirmacher2006Thermal,Schirmacher2007Acoustic} or a spring network~\cite{Wyart2010Scaling,Degiuli2014Effects}, with spatially fluctuating stiffness.
It has succeeded in reproducing several vibrational properties of glasses, such as the non-Debye scaling law near the BP frequency~\cite{Schirmacher2007Acoustic,Degiuli2014Effects}.
In particular, when the theory is applied to jammed systems~\cite{Wyart2010Scaling,During2013Phonon,Degiuli2014Effects,Degiuli2014Force}, it reproduces several power law exponents~\cite{O'Hern2002Random,O'Hern2003Jamming,Silbert2005Vibrations,Silbert2009Normal} by utilizing the marginal stability of amorphous solids.

However, this elasticity theory predicts that the non-Debye scaling becomes gapless when the system is marginally stable\cite{Degiuli2014Effects}.
This is inconsistent with the numerical observations previously indicated.
To reconcile the theory with the numerical data, it has been argued that real glasses are not exactly marginally stable~\cite{Degiuli2014Effects}.
At any rate, however, the theory cannot reproduce the vDOS of the QLVs, $g_{\mathrm{QLV}}(\omega)\sim\omega^\beta$; instead, it illustrates the gapped non-Debye scaling law and the Debye law $g_{\mathrm{Debye}}(\omega)\sim\omega^{d-1}$ in the zero-frequency limit.

The replica theory for the perceptron~\cite{Franz2015Universal} also predicts gapless non-Debye scaling, similar to the elasticity theory.
Recently, a phenomenological attempt was proposed to reproduce the QLVs by introducing spatial fluctuations of stability~\cite{Ikeda2018Note, Ikeda2018Universal}.

All previous studies therefore concluded that real glasses are not exactly marginally stable~\cite{Degiuli2014Effects,Ikeda2018Note,Ikeda2018Universal}.
Namely, we can expect that the theory already captures the nature of the mechanical instability, whereas the remaining task is to identify the parameter region of the phase diagram in which the system is almost marginally stable, as expected.
In contrast, in the preceding study~\cite{Shimada2019Vibrational}, we proposed another mechanism for instability, referred to as local instability, as an alternative to the previous interpretation within the framework of the elasticity theory with quenched disorder.
This new instability corresponds to a local ``defect,'' while the conventional instability that has been analyzed in previous studies occurs when the variance of the disorder distribution is too large.
This overlooked instability is entirely consistent with those of the QLVs, and we presented a toy model that reproduces the gapless quartic law of the vDOS $g(\omega)\sim\omega^4$ when the system is marginally stable by the local instability.
Our results strongly suggest that real glasses are marginally stable, not in the sense of the conventional instability resulting in gapless non-Debye scaling, rather in terms of local instability.

This study presents an extended and thorough analysis of the local instability using the simplest elasticity model.
In contrast to the preceding study~\cite{Shimada2019Vibrational}, several specific examples are presented before the main general argument.
Although the derivation of the local instability is the same as in the preceding study, we can obtain useful insights from those examples.
After introducing the local instability, we present new analytical and numerical calculations, which were not reported in the preceding study.
In particular, we illustrate that when the system is marginally stable, some classes of stiffness distributions yield a gapless vDOS following $g(\omega)\sim\omega^{2\nu+1}$, where $\nu>1$ is an exponent of the distributions.

In Section~\ref{Scalar displacement model}, the scalar displacement model~(SDM)~\cite{Kohler2013Coherent} is analyzed.
In Section~\ref{sec:Model}, the model details are introduced, which is followed by Section~\ref{sec:Approximation} in which an effective medium approximation (EMA) is applied to the model.
In Section~\ref{sec:Large dimension limit}, we discuss the large-dimensional limit of this model in which the EMA becomes exact~\cite{Luck1991Conductivity}.
Conventional instability is introduced in this section.
In Sections~\ref{sec:Uniform distribution} and~\ref{sec:Gaussian distribution}, we investigate specific types of disorders and report contrasting results.
Based on these results, the local instability and related conditions are introduced in Section~\ref{sec:Restrictions on the probability distribution}.
In Section~\ref{sec:Bates distribution at zero frequency}, a series of disorders that illustrate both conventional and local instabilities are presented.
 Finally, in Section~\ref{sec:The lowest-frequency region}, vDOS is derived when the system is marginally stable in terms of the local instability.
The second part, Section~\ref{Vector displacement model}, considers the vector displacement model~(VDM), which has been analyzed in the preceding study~\cite{Shimada2019Vibrational} and in Refs.~\cite{Wyart2010Scaling, During2013Phonon}.
This is nearly equivalent to the model of the first part.
Thus, after introducing the model in Section~\ref{sec:Model2}, we present only the differences from the preceding sections in Section~\ref{sec:Difference from the scalar displacement model}.
Finally, the results are summarized and their implications are analyzed in Section~\ref{sec:Summary and discussion}.

\section{Scalar displacement model}~\label{Scalar displacement model}

\subsection{Model}\label{sec:Model}

To analyze glass vibrations, we first consider the SDM~\cite{Feng1984Percolation,Kohler2013Coherent}.
The model is a $d$-dimensional simple cubic lattice of $N$ elements with unit mass and scalar displacements of $\{u_i\}_{i=1}^N$.
Each nearest-neighbor pair $\left<ij\right>$ is connected by a spring whose stiffness $k_{ij}$ is an independent random variable obeying the probability distribution $P\left(k_{ij}\right)$.

The mean of the distribution, $\mu \equiv \overline{k_{ij}} = \int dk_{ij}k_{ij}P\left(k_{ij}\right)$, must be positive.
Note that our model is considered coarse-grained, and thus the effects of microscopic stress and frustration are encoded as negative stiffness~\cite{Brito2009Geometric,Mizuno2016Spatial}.
Therefore, the negative stiffness is the source of the instability, or, conversely, no instability occurs when all the springs possess positive stiffness.

The equation of motion is given by
\begin{equation}
  \frac{d^2}{dt^2}u_i = -\sum_{j\in\partial i}k_{ij}\left(u_i-u_j\right),
\end{equation}
where $\partial i$ is the set of neighbors of $i$.
Using the bra-ket notation,
\begin{equation}\label{eq:equation of motion}
    \frac{d^2}{dt^2}\ket{u} = -\hat{\mathcal{M}}\ket{u},
\end{equation}
where
\begin{equation}
    \begin{split}
        \hat{\mathcal{M}} &= \sum_{\left<ij\right>}k_{ij}(\ket{i}-\ket{j})(\bra{i}-\bra{j}) \\
        &\equiv \sum_{\alpha=\left<ij\right>}k_\alpha\ket{\alpha}\bra{\alpha}. 
    \end{split}
\end{equation}
is the dynamical matrix.
This is one of the simplest elasticity models.
When the equation of motion is replaced with a master equation, a model for the hopping transport of charge carriers in a disordered semiconductor is obtained~\cite{Kohler2013Coherent}.

 Green's function for Eq.~(\ref{eq:equation of motion}) is defined as $\hat{\mathcal{G}}(\omega) \equiv (\hat{\mathcal{M}}-\omega^2)^{-1}$.
When all springs have the same stiffness $K$, Green's function for the homogeneous system  can be derived as follows:
\begin{equation}
  \begin{split}
    &G_K\left(\boldsymbol{r}_{ij}, \omega\right) \equiv \bra{i}\hat{\mathcal{G}}_K(\omega)\ket{j} \\
    &=\int_{\boldsymbol{q}\in\left[-\pi,\pi\right]^d}\frac{d\boldsymbol{q}}{(2\pi)^{d}}\frac{e^{i\boldsymbol{q}\cdot\boldsymbol{r}_{ij}}}{K\sum_{m=1}^d\left(2-2\cos q_m\right)-\omega^{2}} \\
    &\to \int_{\boldsymbol{q}\in\left[-\pi,\pi\right]^d}\frac{d\boldsymbol{q}}{(2\pi)^{d}}\frac{e^{i\boldsymbol{q}\cdot\boldsymbol{r}_{ij}}}{K\boldsymbol{q}^{2}-\omega^{2}} \qquad (q \ll 1). 
  \end{split}
\end{equation}
where $\boldsymbol{r}_{ij}$ is a vector from the $i$th element to the $j$th element, and the long-wavelength limit is used in the last line.

\subsection{Effective medium approximation}\label{sec:Approximation}

The EMA is introduced in this section, as indicated in Ref.~\cite{Odagaki1981Coherent-medium, Summerfield1981Effective, Webman1981Effective-medium, Feng1985Effective-medium, Wyart2010Scaling, Kohler2013Coherent, Degiuli2014Effects}.
It yields an approximate disorder-averaged Green's function $\overline{\hat{\mathcal{G}}(\omega)}$ within a mean-field-like approach.
The dynamical matrix is decomposed as follows: 
\begin{equation}
  \begin{split}
    \hat{\mathcal{M}} - \omega^2 &= \sum_{\alpha=\langle i j\rangle}k_{\alpha}\ket{\alpha}\bra{\alpha} - \omega^2 \\
    &= \left[k_{\mathrm{eff}}(\omega)\sum_{\alpha=\langle ij \rangle}\ket{\alpha}\bra{\alpha} - \omega^2\right] \\
    &+ \sum_{\alpha=\langle i j\rangle}\left[k_{\alpha}-k_{\mathrm{eff}}(\omega)\right]\ket{\alpha}\bra{\alpha} \\
    &\equiv \hat{\mathcal{G}}_{\mathrm{eff}}(\omega)^{-1} + \hat{\mathcal{V}}(\omega),
  \end{split}
\end{equation}
where $\hat{\mathcal{G}}_{\mathrm{eff}}(\omega) \equiv \hat{\mathcal{G}}_{K=k_{\mathrm{eff}}(\omega)}(\omega)$.
By treating the second term $\hat{\mathcal{V}}(\omega)$ as a perturbation, the transfer matrix can be expressed as follows:
\begin{equation}\label{eq:T matrix series}
  \begin{split}
    \hat{\mathcal{T}}(\omega) &= \sum_{\alpha=\left<ij\right>} \hat{\mathcal{T}}_{\alpha}(\omega) \\&+ \sum_{\alpha=\left<ij\right>} \sum_{\beta\neq\alpha} \hat{\mathcal{T}}_{\alpha}(\omega) \hat{\mathcal{G}}_{\mathrm{eff}}(\omega)\hat{\mathcal{T}}_{\beta}(\omega) + \cdots. 
  \end{split}
\end{equation}
where
\begin{equation}\label{eq:T matrix}
  \hat{\mathcal{T}}_{\alpha}(\omega) = \frac{k_{\mathrm{eff}}(\omega)- k_{\alpha}}{1 - \left[k_{\mathrm{eff}}(\omega)- k_{\alpha}\right] \bra{\alpha}\hat{\mathcal{G}}_{\mathrm{eff}}(\omega)\ket{\alpha}}\ket{\alpha}\bra{\alpha}.
\end{equation}

The self-consistent equation for the effective stiffness $k_{\mathrm{eff}}(\omega)$ is~\cite{Wyart2010Scaling,Degiuli2014Effects}
\begin{equation}
  \overline{\frac{k_{\mathrm{eff}}(\omega)- k_{\alpha}}{ 1 - \left[k_{\mathrm{eff}}(\omega) - k_{\alpha}\right] \bra{\alpha}\hat{\mathcal{G}}_{\mathrm{eff}}(\omega) \ket{\alpha}}} = 0.
\end{equation}
To proceed, we use an identity derived from a trivial relation $\hat{\mathcal{G}}_{\mathrm{eff}}(\omega)\hat{\mathcal{G}}_{\mathrm{eff}}(\omega)^{-1}=\hat{1}$~\cite{Wyart2010Scaling, Degiuli2014Effects}:
\begin{equation}\label{eq:useful identity}
  \bra{\alpha}\hat{\mathcal{G}}_{\mathrm{eff}}\left(\omega\right)\ket{\alpha} = \frac{1}{k_{\mathrm{eff}}d}\left[1 + \omega^2G\left(\omega\right)\right],
\end{equation}
where $G\left(\omega\right) \equiv \bra{i}\hat{\mathcal{G}}_{\mathrm{eff}}(\omega)\ket{i}$.
Thus, the following is obtained:
\begin{equation}\label{eq:self consistent equation}
  \overline{\frac{k_{\mathrm{eff}}(\omega)- k_{\alpha}}{ 1 - \left[k_{\mathrm{eff}}(\omega) - k_{\alpha}\right]\frac{1}{k_{\mathrm{eff}}d}\left[1 + \omega^2G\left(\omega\right)\right] }} = 0,
\end{equation}
where
\begin{equation}\label{eq:Green's function}
  G(\omega) = \int_{\boldsymbol{q}\in\left[-\pi,\pi\right]^d}\frac{d\boldsymbol{q}}{(2\pi)^{d}}\frac{1}{k_{\mathrm{eff}}(\omega)\boldsymbol{q}^{2}-\omega^{2}}.
\end{equation}
In the following, the Debye approximation is applied to Eq.~(\ref{eq:Green's function}); namely, the cubic first Brillouin zone is replaced by a sphere with the same volume
\begin{equation}\label{eq:Green's function2}
  G(\omega) = \int_{0<|\boldsymbol{q}|<q_D} \frac{d\boldsymbol{q}}{(2\pi)^{d}}\frac{1}{k_{\mathrm{eff}}(\omega)\boldsymbol{q}^{2}-\omega^{2}},
\end{equation}
where the radius of the sphere $q_D$ is determined by the condition
\begin{equation}\label{eq:Debye wavenumber}
  1 = \int_{0<|\boldsymbol{q}|<q_D}\frac{d\boldsymbol{q}}{\left(2\pi\right)^d} = \frac{q_D^dS_{d-1}}{d\left(2\pi\right)^d}.
\end{equation}
$S_{d-1}$ is the area of a $(d-1)$-dimensional sphere with radius $1$.

The imaginary part of $G(\omega)$ yields the vDOS as follows:
\begin{equation}
  g(\omega) = \frac{2\omega}{\pi}\Im G(\omega).
\end{equation}
Generally, the effective stiffness is a complex number $k_{\mathrm{eff}}(\omega) = k_r(\omega)-i\Sigma(\omega)$, and when the solution has a finite imaginary part at zero frequency, $\Sigma(0)>0$, the system is unstable.

To solve Eq.~(\ref{eq:self consistent equation}) with some specified $P(k_\alpha)$, it is transformed into a useful form
\begin{equation}\label{eq:self consistent equation with kappa}
  \overline{ \frac{1}{k_\alpha + \kappa(\omega)} } = \frac{1 + \omega^2G\left(\omega\right)}{dk_{\mathrm{eff}}(\omega)},
\end{equation}
where
\begin{equation}\label{eq:definition of kappa}
  \kappa(\omega) \equiv \frac { d -  1 - \omega ^ { 2 } { G } ( \omega )} { 1 + \omega ^ { 2 } { G } ( \omega ) } k _ { \mathrm{eff} }(\omega).
\end{equation}
Note that $\kappa(\omega)\to (d-1)k_{\mathrm{eff}}(0)$ as $\omega\to0$ and that $\Im\kappa(\omega)\sim\Im k_{\mathrm{eff}}(\omega)<0$ can be assumed if we focus on the low-frequency region.

\subsection{Large dimension limit}\label{sec:Large dimension limit}

In this section, Eq.~(\ref{eq:self consistent equation}) is solved in the large-dimension limit.
This is crucial because the EMA becomes exact as $d\to\infty$~\cite{Luck1991Conductivity}.
We do not assume the specific form of the distribution $P(k_\alpha)$ and only require that its moment-generating function is finite.

As $d\to\infty$ and $\omega\to0$, $G(\omega)$ in Eq.~(\ref{eq:Green's function2}) can be approximated as follows:
\begin{equation}\label{eq:approximate Green's function for large dimension}
  G(\omega) \simeq \frac{1}{k_{\mathrm{eff}}(\omega)q_D^2 - \omega^2}.
\end{equation}
Its derivation is provided in Appendix~\ref{sec:Approximation of the Green's function in the large dimension limit}.
Using Eq.~(\ref{eq:approximate Green's function for large dimension}), Eq.~(\ref{eq:self consistent equation}) becomes
\begin{equation}
  \overline{\frac{k_{\mathrm{eff}} - k_{\alpha}}{ 1 - \frac{k_{\mathrm{eff}}(\omega)-k_{\alpha}}{d\left[k_{\mathrm{eff}}(\omega) - \omega^2/q_D^2\right]}}} = 0.
\end{equation}
In the large-dimension limit, the denominator can be expanded as follows ~\cite{Kohler2013Coherent}:
\begin{equation}
  k_{\mathrm{eff}}(\omega) - \overline{k_{\alpha}} + \frac{1}{d}\frac{\overline{\left[k_{\mathrm{eff}}(\omega)-k_{\alpha}\right]^2}}{k_{\mathrm{eff}}(\omega) - \omega^2/q_D^2} = 0.
\end{equation}
Therefore, the self-consistent equation is expressed as follows:
\begin{widetext}
\begin{equation}\label{eq:infinite dimensional equation}
    \begin{split}
        0 & = (1+1/d) k_{\mathrm{eff}}(\omega)^2 - [ (1+2/d)\mu + \omega^2/q_D^2 ] k_{\mathrm{eff}}(\omega) + \mu\omega^2/q_D^2 + (\sigma^2 + \mu^2)/d \\
        &\simeq k_{\mathrm{eff}}(\omega)^2 - (\mu + \omega^2/q_D^2)k_{\mathrm{eff}}(\omega) + \mu\omega^2/q_D^2 + (\sigma^2 + \mu^2)/d = 0, 
    \end{split}
\end{equation}
where $\sigma^2$ is the variance of the distribution $P(k_\alpha)$; $\sigma^2 \equiv \overline{(k_\alpha-\mu)^2} = \int dk_\alpha(k_{\alpha}-\mu)^2P\left(k_\alpha\right)$, and we neglected unimportant terms of order $\mathcal{O}(d^{-1})$ that do not change the solution qualitatively.
Two solutions are obtained by solving this quadratic equation, and the one that satisfies $k_{\mathrm{eff}} = \mu$ at $\sigma = \omega = 0$ is chosen:
\begin{equation}\label{eq:infinite dimensional solution}
    k_{\mathrm{eff}}(\omega) = \frac{1}{2}(\mu + \omega^2/q_D^2) + \frac{1}{2}\sqrt{\mu^2 - 4\sigma^2/d - 2\mu\omega^2/q_D^2 + \omega^4/q_D^4},
\end{equation}
\end{widetext}
where small terms in the large-dimension limit are neglected again.

The critical value of the standard deviation $\sigma_c$, above which the system is unstable, $\Sigma(0)>0$, is $\sigma_c = \sqrt{d}\mu/2 \sim d^{1/2}\mu$.
Its dependence on dimension is simple to understand as follows.
The mean and the standard deviation of the sum of all spring constants attached to an element are of order $d\mu$ and $d^{1/2}\sigma$, respectively.
When they are of the same order, which gives $\sigma_c$, the system is destabilized.
This instability has been observed in several elasticity models with perturbations~\cite{Taraskin2003Vector,Schirmacher2007Acoustic,Degiuli2014Force} and we call it the conventional instability~\cite{Shimada2019Vibrational}.

When $\sigma\leq\sigma_c$, a quantity which measures the distance to the critical value can be defined
\begin{equation}
  \omega_0 \equiv q_D\sqrt{\left(\mu^2-4\sigma^2/d\right)/2\mu} = \sqrt{\frac{2q_D^2}{d\mu}}\sqrt{\sigma_c^2-\sigma^2}.
\end{equation}
Using this frequency, Eq.~(\ref{eq:infinite dimensional solution}) can be expressed as
\begin{equation}\label{eq:infinite dimensional solution omega_0}
  k_{\mathrm{eff}}(\omega) = \frac{1}{2}(\mu + \omega^2/q_D^2) + \sqrt{(\mu/2q_D^2)(\omega_0^2 - \omega^2) + \omega^4/q_D^4}.
\end{equation}
Therefore, when we focus on the low-frequency region and neglect the term of order $\mathcal{O}(\omega^4)$, the behavior of this effective stiffness changes at $\omega = \omega_0$.
When $\omega<\omega_0$, we have $k_r(\omega) = k_{\mathrm{eff}}(\omega)$ and $\Sigma(\omega) = 0$.
Expanding the real part near $\omega=0$, we obtain 
\begin{equation}
    k_r(\omega) = \frac{\mu}{2} + \sqrt{\frac{\mu\omega_0^2}{2q_D^2}} - \frac{\omega_0^2}{2q_D^2}\left( \sqrt{\frac{1}{1-\sigma^2/\sigma_c^2}} - 1\right) + \mathcal{O}(\omega^4).
\end{equation}
Thus, the real part decreases in this frequency region.
However, the imaginary part is always zero.
This is an artifact of our approximation in which the pole of Green's function responsible for the Rayleigh scattering is neglected ~\cite{Degiuli2014Force}.
However, even when we incorporate the pole, it only yields the vanishingly small imaginary part $\Sigma(\omega)\sim\omega^d$ in the low-frequency region~\cite{Degiuli2014Force}.
By contrast, $k_r(\omega) = \mu/2 + \omega^2/2q_D^2$ and $\Sigma(\omega) = \sqrt{(\mu/2q_D^2)(\omega^2 - \omega_0^2)}$ when $\omega_0<\omega$.
Thus, the real part starts to increase, which leads to a local minimum in the phase velocity of sound near $\omega=\omega_0$, often called sound softening~\cite{Wyart2010Scaling}.
The imaginary part follows a power law $\Sigma(\omega)\sim\omega$ when $\omega\gg\omega_0$.

The vDOS is given by
\begin{equation}
  g\left(\omega\right) = \frac{2\omega}{\pi}\Im G\left(\omega\right) \sim \omega\Sigma.
\end{equation}
Therefore, $g\left(\omega\right)\sim\omega^2$ when $\omega\gg\omega_0$, which is called the non-Debye scaling~\cite{Degiuli2014Effects, Franz2015Universal, Charbonneau2016Universal}.
As $\sigma\to\sigma_c$, we have $\omega_0\to0$, which gives the gapless non-Debye scaling.
Note that, in the SDM, the non-Debye scaling is universal among any distribution in the large-dimension limit if the distribution $P(k_\alpha)$ has finite moments.

\subsection{Uniform distribution of stiffness}\label{sec:Uniform distribution}

We now focus on specific distributions $P(k_\alpha)$ in finite dimensions under the EMA.
Note that the EMA is not only exact as $d\to\infty$ but is also a good approximation in finite $d$~\cite{Luck1991Conductivity}.
First, we consider the model with a uniform distribution
\begin{equation}
  P(k_\alpha) =
  \begin{cases}
    \frac{1}{2\Delta} & k_\alpha\in[\mu-\Delta, \mu+\Delta] \\
    0 & \mathrm{otherwise}
  \end{cases}
  .
\end{equation}
The variance is $\sigma^2 = \Delta^2/3$.
This model provides essentially the same results as those in the large-dimension limit. 
The self-consistent equation with a uniform distribution is easy to solve and has been used in previous studies~\cite{Taraskin2003Vector,Kohler2013Coherent}.
Although our analysis is almost equivalent to the previous studies, this model is a useful example for the following discussion. Therefore, we present the results for completeness.

By averaging Eq.~(\ref{eq:self consistent equation with kappa}) over $k_\alpha$, we obtain
\begin{equation}\label{eq:expansion of the self consistent equation}
  \begin{split}
    \frac{\mu + \kappa(\omega)}{\Delta} &= \coth\left\{\frac{\Delta}{dk_{\mathrm{eff}}(\omega)}\left[1 + \omega^2G(\omega)\right]\right\} \\
    &\simeq \frac{dk_{\mathrm{eff}}(\omega)}{\Delta\left[1 + \omega^2G(\omega)\right]} + \frac{1}{3}\frac{\Delta}{dk_{\mathrm{eff}}(\omega)}\left[1 + \omega^2G(\omega)\right].
  \end{split}
\end{equation}
In the second line, a series expansion of $\coth$ is used, which is justified when $\Delta/d\mu\ll1$.
Using the definition of $\kappa(\omega)$ in Eq.~(\ref{eq:definition of kappa}), the equation is simplified to
\begin{equation}\label{eq:self consistent equation for the uniform distribution}
  k_{\mathrm{eff}}(\omega)^2 -\mu k_{\mathrm{eff}}(\omega) + \frac{\sigma^2}{d} + \frac{\sigma^2}{d}\omega^2G(\omega) = 0.
\end{equation}
We focus on the lowest-frequency region and approximate Green's function in Eq.~(\ref{eq:Green's function2}) as follows: 
\begin{equation}\label{eq:approximate Green's function}
  \begin{split}
    G(\omega) &\simeq \frac{1}{k_{\mathrm{eff}}(\omega)}\int_{0<|\boldsymbol{q}|<q_D}\frac{d\boldsymbol{q}}{(2\pi)^d}\frac{1}{\boldsymbol{q}^2} \\
    &= \frac{S_{d-1}}{k_{\mathrm{eff}}(\omega)(2\pi)^d}\int_0^{q_D}dq q^{d-3} \\
    &= \frac{1}{k_{\mathrm{eff}}(\omega)}\frac{d}{(d-2)q_D^2} \equiv \frac{A_d}{k_{\mathrm{eff}}(\omega)},
  \end{split}
\end{equation}
where the third line is obtained using Eq.~(\ref{eq:Debye wavenumber}). 
Therefore, Eq.~(\ref{eq:self consistent equation for the uniform distribution}) reduces to a cubic equation
\begin{equation}\label{eq:self consistent equation for uniform distribution}
  k_{\mathrm{eff}}(\omega)^3 -\mu k_{\mathrm{eff}}(\omega)^2 + \frac{\sigma^2}{d}k_{\mathrm{eff}}(\omega) + \frac{\sigma^2}{d}A_d\omega^2 = 0.
\end{equation}

At zero frequency, this equation is the same as Eq.~(\ref{eq:infinite dimensional equation}), and $\sigma = \sigma_c \equiv \sqrt{d}\mu/2$ is the critical point for stability.
When $\omega\neq0$, the equation is approximated under the condition $(\sigma_c^2-\sigma^2)/d\mu^2 \sim \omega^2/\mu \ll 1$ in Appendix~\ref{sec:Solution for the uniform distribution}, which yields
\begin{equation}
  k_{\mathrm{eff}}(\omega) = \frac{\mu}{2} - i\sqrt{\frac{\mu}{2}}\sqrt{A_d\omega^2-A_d{\omega'_{0}}^2},
\end{equation}
where
\begin{equation}
  \omega_0' \equiv \sqrt{\frac{2}{d\mu A_d}}\sqrt{\sigma_c^2-\sigma^2}.
\end{equation}
Note that $A_d\to q_D^{-2}$ and $\omega_0'\to\omega_0$ as $d\to\infty$.
This is consistent with Eq.~(\ref{eq:infinite dimensional solution omega_0}).
For a uniform distribution, the solution in the large-dimension limit is a sufficient approximation, even in finite dimensions.
We can confirm that the error of the approximation in Eq.~(\ref{eq:expansion of the self consistent equation}) is significantly small in $d=3$.
The SDM with a uniform distribution is destabilized by the conventional instability.

\begin{figure*}[t]
  \begin{center}
    \includegraphics[width=0.9\textwidth]{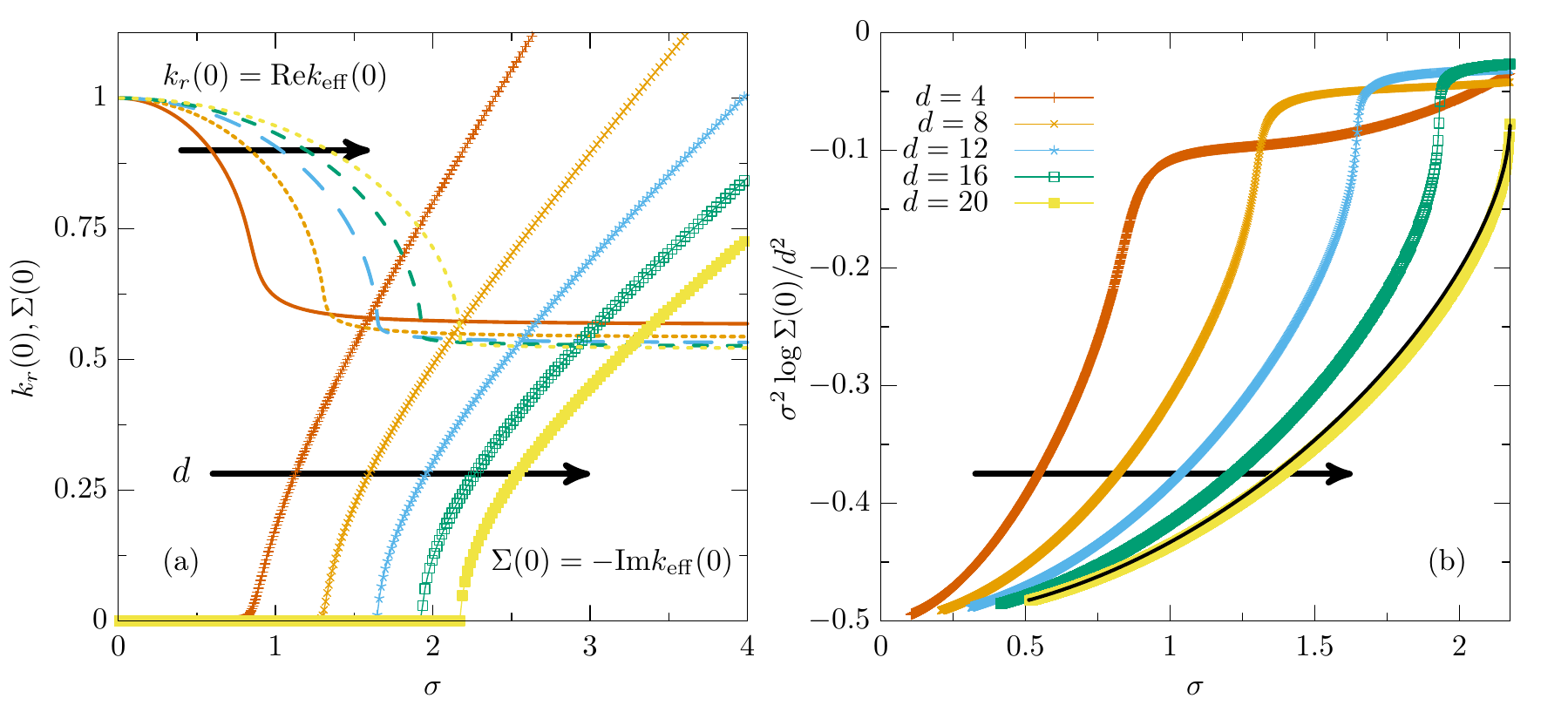}
    \vspace*{0 mm}
    \caption{
      (a) Real and imaginary parts of the effective stiffness for the model with the Gaussian distribution as functions of $\sigma$ at zero frequency.
      The spatial dimension is changed from $d=4$ to $d=20$ (left to right curves).
      We set $\mu=1$.
      (b) $\sigma^2\log\Sigma(0)$ vs $\sigma$.
      The solid line indicates the approximate form of the large-dimension limit in Eq.~(\ref{eq:Gaussian imaginary}) in which we substitute $d=20$.
    }
    \label{fig:fig1}
  \end{center}
\end{figure*}

\begin{figure*}[t]
  \begin{center}
    \includegraphics[width=0.9\textwidth]{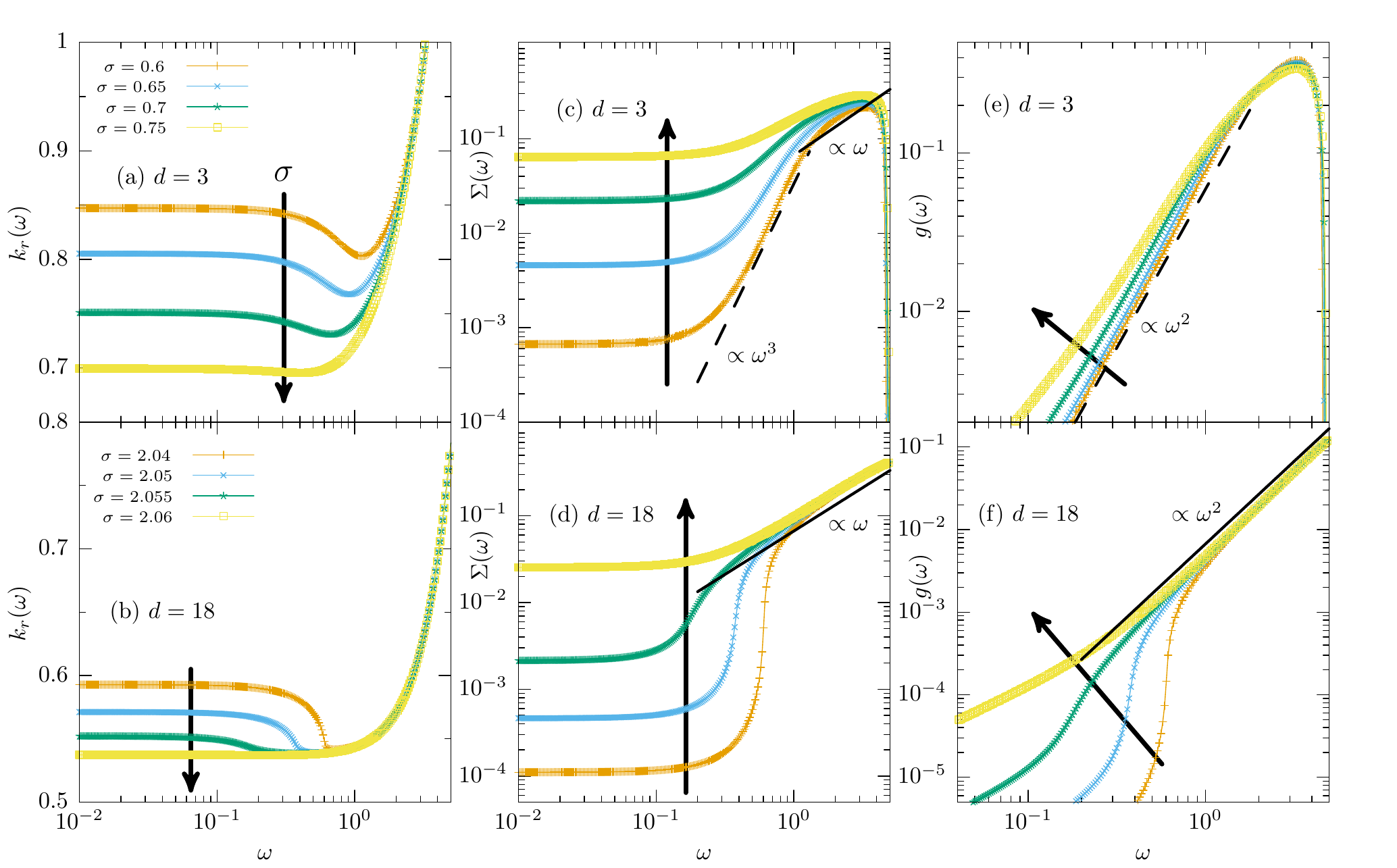}
    \vspace*{0 mm}
    \caption{
      (a) Real parts of effective stiffness as functions of frequency for $\sigma$ in $d=3$ .
      $\sigma$ is changed from $\sigma=0.6$ to $\sigma=0.75$ (top to bottom curves).
      (b) The same plot as the panel (a) in $d=18$.
      $\sigma$ is changed from $\sigma=2.04$ to $\sigma=2.06$ (top to bottom curves).
      (c) Imaginary parts of effective stiffness as functions of frequency in $d=3$.
      Values of $\sigma$ are the same as in the panel (a).
      The solid line is proportional to $\omega$, and the dashed line is proportional to $\omega^d=\omega^3$.
      (d) The same plot as the panel (c) in $d=18$.
      Values of $\sigma$ are the same as in the panel (b).
      (e) vDOS in $d=3$ for the same values of $\sigma$ as in the panel (a).
      The tail of vDOS represents $g(\omega)\sim\omega^2$ (dashed line), which is the Debye scaling of phonons.
      (f) The same plot as in the panel (e) in $d=18$ for the same values of $\sigma$ as in the panel (b).
      The Debye scaling $\omega^{d-1}=\omega^{17}$ is practically impossible to observe, whereas the non-Debye scaling $g(\omega)\sim\omega^2$ (solid line) grows compared to the $d=3$ case.
    }
    \label{fig:fig2}
  \end{center}
\end{figure*}

\subsection{Gaussian distribution of stiffness}\label{sec:Gaussian distribution}

Next, a Gaussian distribution with a mean of $\mu$ and variance of $\sigma^2$ is considered. 
This seems natural for coarse-grained stiffness~\cite{Schirmacher2006Thermal,Mizuno2016Spatial}.
The Gaussian distribution produces results that are qualitatively different from those of the uniform distribution.
The results of the Gaussian distribution are key to understanding the mechanism behind the stability of the system within the framework of the EMA.

By averaging Eq.~(\ref{eq:self consistent equation with kappa}) with the Gaussian distribution, we obtain 
\begin{equation}
  \begin{split}
    \frac{1}{\sqrt{2\pi}}\int_{-\infty}^\infty dx\frac{e^{-x^2}}{x+z} = \frac{\sigma}{dk_{\mathrm{eff}}(\omega)}\left[1 + \omega^2G\left(\omega\right)\right], \\
  \end{split}
\end{equation}
where $\Im\kappa(\omega)<0$ and $z = [\mu + \kappa(\omega)]/\sqrt{2\sigma^2}$.
The left-hand side is further calculated as follows:
\begin{equation}
  \begin{split}
    \frac{1}{\sqrt{2\pi}}\int_{-\infty}^\infty dx\frac{e^{-x^2}}{x+z} &= \sqrt{2}e^{-z^2}\int_0^z dl e^{l^2} + i\sqrt{\frac{\pi}{2}}e^{-z^2}.
  \end{split}
\end{equation}
Therefore, the self-consistent equation is expressed as follows:
\begin{equation}\label{eq:self consistent equation for the Gaussian distribution}
  F(z) + i\frac{\sqrt{\pi}}{2}e^{-z^2} = \frac{\sigma}{\sqrt{2}dk_{\mathrm{eff}}(\omega)}\left[1 + \omega^2G\left(\omega\right)\right],
\end{equation}
where $F(z)$ is the Dawson function.

\subsubsection{Zero-frequency limit}\label{sec:Zero frequency}

It is instructive to consider the zero-frequency limit.
In this case, Eq.~(\ref{eq:self consistent equation for the Gaussian distribution}) becomes 
\begin{widetext}
\begin{equation}\label{eq:gaussian self consistent zero frequency}
  F\left[\frac{\mu + (d-1)k_{\mathrm{eff}}(0)}{\sqrt{2\sigma^2}}\right] + i\frac{\sqrt{\pi}}{2}\exp\left\{-\left[\frac{\mu + (d-1)k_{\mathrm{eff}}(0)}{\sqrt{2\sigma^2}}\right]^2\right\} = \frac{\sigma}{\sqrt{2}dk_{\mathrm{eff}}(0)}.
\end{equation}
\end{widetext}
Note that if the imaginary part of $k_{\mathrm{eff}}(0)$ is zero (or infinitesimally small), the equation does not hold because of the second term on the left-hand side.
That is, there cannot be a stable solution at any $\sigma$ for the Gaussian distribution.
This is somewhat counter-intuitive, which is further discussed in Section~\ref{sec:Restrictions on the probability distribution}.

The instability of the Gaussian distribution can be observed in any dimension.
However, as indicated in Section~\ref{sec:Large dimension limit}, a stable solution can be obtained when $\sigma<\sigma_c=\sqrt{d}\mu/2$ in the large-dimension limit.
Here, we present how the unstable solution for the Gaussian distribution asymptotically converges to the stable solution in Eq.~(\ref{eq:infinite dimensional solution}) as $d\to\infty$.
Because the calculation is straightforward but tedious, we present it in Appendix~\ref{sec:Asymptotic solution for the Gaussian distribution in the large dimension limit} and only present the results here.
As $d\to\infty$, the real part is
\begin{equation}\label{eq:Gaussian real}
  k_r(0) = \frac { \mu } { 2 } + \frac { 1 } { 2 } \sqrt { \mu ^ { 2 } - 4 \frac { \sigma ^ { 2 } } { d } }
\end{equation}
and the imaginary part is
\begin{equation}\label{eq:Gaussian imaginary}
   \Sigma(0) = \sqrt { \frac { \pi } { 2 } } \frac { d ^ { 2 } k_r(0) ^ { 4 } \exp \left[ - \frac { d ^ { 2 } k_r(0) ^ { 2 } } { 2 \sigma ^ { 2 } } \right]} { \mu \sigma \left[k_r(0) - 2 \frac { \sigma ^ { 2 } } { \mu d } \right]} .
\end{equation}
When $\sigma = \sqrt{d}(\mu-\epsilon)/2$ is set with $0<\epsilon\ll1$, the real part becomes $k_r(0) = { \mu } / { 2 } + \sqrt { 2 \mu \epsilon }$.
Therefore, the imaginary part is expressed as follows:
\begin{equation}\label{eq:singularity of Gaussian}
  \Sigma(0) \simeq \frac { \sqrt { \pi } } { 16 } \frac { d ^ { 3 / 2 } \mu ^ { 2 } } { \sqrt { \mu \epsilon } } e ^ { - d / 2 } \sim d ^ { 3 / 2 } e ^ { - d / 2 } \epsilon ^ { - 1 / 2 }.
\end{equation}
This is exponentially small when $\sigma\ll\sqrt{d}\mu/2$ and asymptotically vanishes as $d\to\infty$.
Therefore, as $d\to\infty$, the solution converges to Eq.~(\ref{eq:infinite dimensional solution}).
The imaginary part grows rapidly when $\sigma$ approaches the critical value $\sigma_c\equiv\sqrt{d}\mu/2$ and exhibits singular behavior $\Sigma(0)\sim \epsilon ^ { - 1 / 2 }$, which can be interpreted as a transition in the large-dimension limit.

We also numerically solved Eq.~(\ref{eq:gaussian self consistent zero frequency}), and the results are presented in Fig.~\ref{fig:fig1}.
In this computation, we set $\mu=1$.
Figure~\ref{fig:fig1}(a) shows the effective stiffness as functions of the standard deviation in $d=4,\ 8,\ 12,\ 16$, and $20$.
When $\sigma=0$, a trivial solution is obtained: $k_r(0)=\mu=1$ and $\Sigma(0)=0$.
When we increase $\sigma$, the real part starts to decrease and becomes almost flat at a certain value of $\sigma$ depending on the spatial dimension.
The imaginary part appears to be zero in the small-$\sigma$ region, but it is nonzero, as shown in Fig.~\ref{fig:fig1}(b).
It starts to grow rapidly when the real part becomes flat.
This crossover value of $\sigma$ converges to the transition value $\sigma_c=\sqrt{d}\mu/2$ in the large-dimension limit.
Figure~\ref{fig:fig1}(b) presents $\sigma^2\log\Sigma(0)/d^2$ vs $\sigma$.
As indicated above, the imaginary part is always nonzero when $\sigma\neq0$.
We also plot the approximate form of the large-dimension limit in Eq.~(\ref{eq:Gaussian imaginary}), in which we substitute $d=20$, by the solid line.
The sufficient agreement between the numerical solution and Eq.~(\ref{eq:Gaussian imaginary}) indicates that the solution in $d=20$ can be approximated by the one in the large-dimension limit.

\subsubsection{Finite frequency}\label{sec:Finite frequency}

For finite frequency, Eq.~(\ref{eq:self consistent equation for the Gaussian distribution}) is numerically solved, and the results are presented in Fig.~\ref{fig:fig2}.
Figure~\ref{fig:fig2}(a) depicts the real parts of the effective stiffness $k_r(\omega)=\Re k_{\mathrm{eff}}(\omega)$ for $\sigma=0.6,\ 0.65,\ 0.7,$ and $0.75$ in $d=3$.
They present local minima at approximately $\omega=1$, which are also discussed in Section~\ref{sec:Large dimension limit}~\cite{Wyart2010Scaling}.
Here, we denote the position of the minimum by $\omega_{\mathrm{min}}$.
When $\sigma$ is increased, the minimum gradually becomes vague and $\omega_{\mathrm{min}}$ decreases.
Figure~\ref{fig:fig2}(b) presents the equivalent results in $d=18$ for $\sigma=2.04,\ 2.05,\ 2.065,$ and $2.06$.
The qualitative behavior is the same as that shown in Fig.~\ref{fig:fig2}(a), but the change of the local minimum is significantly sharper within this narrower range of $\sigma$ than in $d=3$.
$\omega_{\mathrm{min}}$ decreases rapidly and the minimum almost disappears at $\sigma=2.06$.
As $d\to\infty$, $\omega_{\mathrm{min}}$ converges to $\omega_0$ defined in Section~\ref{sec:Large dimension limit}.
Thus, the disappearance of the local minimum corresponds to the instability even in the large-dimension limit, i.e., the conventional instability although the finite-dimensional model with the Gaussian distribution is always unstable in the sense of the local instability as discussed in Section~\ref{sec:Restrictions on the probability distribution} shortly.

Figure~\ref{fig:fig2}(c) depicts the imaginary parts $\Sigma(\omega)=-\Im k_{\mathrm{eff}}(\omega)$ in $d=3$ for the same values of $\sigma$, as shown in Fig.~\ref{fig:fig2}(a).
In the lowest-frequency region, they converge to the values in the zero-frequency limit.
When the frequency is increased, we obtain the scaling $\Sigma(\omega)\sim\omega^d$, which is characteristic of the Rayleigh scattering~\cite{Wyart2010Scaling, Degiuli2014Effects}.
This scaling is apparent for the smallest value of $\sigma$, but is smeared when $\sigma$ is increased.
In the highest-frequency region, the non-Debye scaling $\Sigma(\omega)\sim\omega$ can be observed, which was indicated in Section~\ref{sec:Large dimension limit} (see Eq.~(\ref{eq:infinite dimensional solution omega_0})).
Figure~\ref{fig:fig2}(d) is the equivalent plot in $d=18$ for the same values of $\sigma$, as shown in Fig.~\ref{fig:fig2}(b).
It is difficult to observe the contribution from the Rayleigh scattering $\Sigma(\omega)\sim\omega^d$; instead, the non-Debye scaling $\Sigma(\omega)\sim\omega$ region grows significantly compared to that in Fig~\ref{fig:fig2}(c).

Figures~\ref{fig:fig2}(e) and (f) depict the corresponding vDOS in $d=3$ and $=18$, respectively.
Although the same frequency dependence $g(\omega)\sim\omega^2$ is evident in both plots, their meanings are different from each other.
In Fig.~\ref{fig:fig2}(c), it is the Debye scaling of phonons $g_{\mathrm{Debye}}(\omega)\sim\omega^{d-1}$, whereas in Fig.~\ref{fig:fig2}(d), it is the non-Debye scaling, which implies quadratic frequency dependence regardless of $d$.
The ranges of the non-Debye scaling in $d=3$ and the Debye scaling in $d=18$ are too narrow to observe.
Note that the linear frequency dependence in the lowest-frequency region, which is particularly evident in Fig.~\ref{fig:fig2}(f), is simply caused by the plateau of the zero-frequency value of $\Sigma(\omega)$.

\subsection{Restrictions on probability distribution}\label{sec:Restrictions on the probability distribution}

\subsubsection{Local instability}~\label{sec:Distributions supported on the whole real line}

In this section, we consider why the Gaussian distribution cannot provide a stable solution from general arguments.
Another instability mechanism is introduced here, which is referred to as the ``local instability.''
The argument is the same as in the preceding study~\cite{Shimada2019Vibrational}; however, it should be easier to understand with the examples provided in the preceding sections.

Because only the stability of the system is of interest, it is sufficient to consider the zero-frequency limit.
Thus, we consider the self-consistent equation Eq.~(\ref{eq:self consistent equation with kappa}) in the zero-frequency limit
\begin{equation}\label{eq:zero frequency self consistent equation}
  \int\frac{dk_\alpha P(k_\alpha)}{k_\alpha + (d-1)k_{\mathrm{eff}}(0)} = \frac{1}{dk_{\mathrm{eff}}(0)}.
\end{equation}
Suppose that the solution has only an infinitesimally small imaginary part, that is, $k_{\mathrm{eff}}(0)-i\epsilon$ with $\epsilon \ll 1$.
Therefore, the left-hand side of Eq.~(\ref{eq:zero frequency self consistent equation}) becomes
\begin{equation}
  \mathcal{P}\int\frac{dk_\alpha P(k_\alpha)}{k_\alpha + (d-1)k_{\mathrm{eff}}(0)} + i\pi P\left[-\left(d-1\right)k_{\mathrm{eff}}(0)\right],
\end{equation}
where $\mathcal{P}$ indicates the Cauchy principal value.
Therefore, the condition
\begin{equation}\label{eq:stability condition}
  P\left[-\left(d-1\right)k_{\mathrm{eff}}(0)\right] = 0
\end{equation}
is necessary for the solution to be real.
Conversely, if this condition is violated, the system is unstable with the emergence of the imaginary part, $\Sigma(0)>0$.
This is the origin of local instability.
This readily leads to the fact that for all distributions which are nonzero on the entire real line $\mathbb{R}$, e.g., the Gaussian distribution cannot yield a stable solution.

\subsubsection{Interpretation of local instability using defect model}

A defect model is considered to illustrate a simple physical interpretation of the local instability. In this model, all springs have the same stiffness of $k_{\mathrm{eff}}(0)>0$, except for a defect with stiffness $k_\alpha$. 

The dynamical matrix of the defect model is 
\begin{equation}\label{M of defect model}
    \begin{split}
        \hat{\mathcal{M}}_d &= k_{\mathrm{eff}}(0)\sum_{\beta=\langle k l\rangle}\ket{\beta}\bra{\beta} + \left[k_{\alpha}-k_{\mathrm{eff}}(0)\right]\ket{\alpha}\bra{\alpha}.
    \end{split}
\end{equation}
Upon calculating the transfer matrix for this dynamical matrix, only the first term of the expansion in Eq.~(\ref{eq:T matrix series}) remains, and the total Green's function is 
\begin{equation}
    \hat{\mathcal{G}}_d(\omega) = \hat{\mathcal{G}}_0(\omega) + \hat{\mathcal{G}}_0(\omega)\hat{\mathcal{T}}_{\alpha0}(\omega)\hat{\mathcal{G}}_0(\omega),
\end{equation}
where $\hat{\mathcal{G}}_0(\omega) = \hat{\mathcal{G}}_{K=k_{\mathrm{eff}}(0)}(\omega)$, and $\hat{\mathcal{T}}_{\alpha0}(\omega)$ is expressed by Eq.~(\ref{eq:T matrix}) with $k_{\mathrm{eff}}(\omega)\to k_{\mathrm{eff}}(0)$ and $\hat{\mathcal{G}}_{\mathrm{eff}}(\omega)\to\hat{\mathcal{G}}_{0}(\omega)$, that is,
\begin{equation}\label{eq:T matrix2}
  \hat{\mathcal{T}}_{\alpha0}(\omega) = \frac{k_{\mathrm{eff}}(0)- k_{\alpha}}{1 - \left[k_{\mathrm{eff}}(0)- k_{\alpha}\right] \bra{\alpha}\hat{\mathcal{G}}_{0}(\omega)\ket{\alpha}}\ket{\alpha}\bra{\alpha}.
\end{equation}
From Eq.~(\ref{eq:T matrix2}), $\hat{\mathcal{T}}_{\alpha0}(\omega)$ diverges when $k_\alpha=-(d-1)k_{\mathrm{eff}}(0)$, which indicates that the system has a nontrivial zero mode.

This zero mode is given by (without normalization)
\begin{equation}
    \ket{0} \equiv \hat{\mathcal{G}}_0(0)\ket{\alpha}.
\end{equation}
The eigenrelation $\hat{\mathcal{M}}\ket{0} = 0$ can be checked as follows.
The product of the first term in Eq.~(\ref{M of defect model}) and $\ket{0}$ yields $\ket{\alpha}$.
The second term yields 
\begin{equation}
    \begin{split}
        &\left[k_{\alpha}-k_{\mathrm{eff}}(0)\right]\ket{\alpha}\braket{\alpha|0} \\
        &= \left[k_{\alpha}-k_{\mathrm{eff}}(0)\right]\ket{\alpha}\bra{\alpha}\hat{\mathcal{G}}_0(0)\ket{\alpha} \\
        &= \frac{\theta}{k_{\mathrm{eff}}}\left[k_{\alpha}-k_{\mathrm{eff}}(0)\right]\ket{\alpha},
    \end{split}
\end{equation}
where we use Eq.~(\ref{eq:useful identity}). 
Therefore, we obtain 
\begin{equation}\label{M0}
    \hat{\mathcal{M}}\ket{0} = \left\{ 1 + \frac{\theta [k_\alpha - k_{\mathrm{eff}}(0)]}{k_{\mathrm{eff}}(0)} \right\} \ket{\alpha}.
\end{equation}
Thus, $\hat{\mathcal{M}}\ket{0} = 0$ when the ``defect'' bond $\alpha$ has a negative stiffness $k_\alpha = -(d-1)k_{\mathrm{eff}}(0)$. 

When $k_\alpha < -(d-1)k_{\mathrm{eff}}(0)$, the system is unstable along the direction $\ket{0}$. 
This provides an interpretation of the stability condition Eq.~(\ref{eq:stability condition}); when it is violated, a number of springs become ``defects'' and produce unstable modes. 
Note that the instability identified here is different from the conventional instability indicated in the preceding sections.
Therefore, the singular behavior of the solution for the Gaussian distribution (see Eq.~(\ref{eq:singularity of Gaussian})) can be interpreted as a transition from the local instability to the conventional instability.

The unstable modes associated with the local instability are significantly similar to the QLVs.
First, the mode $\ket{0}$ is the response to a local dipolar force in an unperturbed homogeneous system.
The elasticity theory illustrates that this response field has an asymptotic spatial profile $\propto r^{2(1-d)}$, where $r$ is the distance to the force, and the QLVs have the same profile far from the core~\cite{Lerner2014Breakdown,Lerner2016Statistics}.
Second, the response to the dipolar force in glasses has a core whose size is the same as that of the QLVs~\cite{Yan2016On,Shimada2018Spatial}.
Third, the energetics of the QLVs are equivalent to $\ket{0}$; the unstable core corresponds to the second term in Eq.~(\ref{M0}), whereas the stable far-field components correspond to the first term in Eq.~(\ref{M0}).

\subsubsection{Restriction on tail to avoid local instability}\label{sec:Restrictions on the tail}

Generally, we do not know which instability is caused by a particular distribution, but can derive a sufficient condition to avoid the local instability.
Namely, it is necessary to violate the condition derived in this section to cause the local instability.
We consider a distribution which is nonzero only on a finite interval $[-k_{\mathrm{min}}, k_{\mathrm{max}}]$ with $k_{\mathrm{min}},k_{\mathrm{max}} > 0$.
If the distribution is assumed to yield the solution $k_{\mathrm{eff}}(0) = k_{\mathrm{min}}/(d-1)$, which marginally satisfies the condition in Eq.~(\ref{eq:stability condition}), then from Eq.~(\ref{eq:zero frequency self consistent equation}), the equation :
\begin{equation}\label{eq:critical self consistent equation}
  \int_{-k_{\mathrm{min}}}^{k_\mathrm{max}}\frac{dk_\alpha P(k_\alpha)}{k_\alpha + k_{\mathrm{min}}} = \frac{d-1}{dk_{\mathrm{min}}}
\end{equation}
holds at zero frequency.
The condition that avoids the local instability can be derived by considering the finite-frequency solution of this distribution.
Because only the lowest-frequency region is considered, we use Eq.~(\ref{eq:approximate Green's function}) for Green's function, and hence, can approximate $\kappa(\omega)$ as
\begin{equation}
  \begin{split}
    \kappa(\omega) &= (d-1)k_r(\omega) - dA_d\omega^2 - i(d-1)\Sigma(\omega).
  \end{split}
\end{equation}
Furthermore, since $k_r(0) = \mathcal{O}(1)$ and $\Sigma(0)=0$, we can assume that $\Re \kappa(\omega)\gg\Im \kappa(\omega)$ when $\omega$ is sufficiently small.
Thus, the self-consistent equation in Eq.~(\ref{eq:self consistent equation with kappa}) for the real part is
\begin{widetext}
\begin{equation}\label{eq:equation for the real part}
  \mathcal{P}\int_{-k_{\mathrm{min}}}^{k_{\mathrm{max}}}\frac{dk_\alpha P(k_{\alpha})}{k_\alpha + (d-1)k_r(\omega) - dA_d\omega^2} = \frac{1}{dk_r(\omega)}\left[1 + \frac{A_d\omega^2}{k_r(\omega)}\right].
\end{equation}
Subtracting Eq.~(\ref{eq:critical self consistent equation}) from Eq.~(\ref{eq:equation for the real part}), we obtain
\begin{equation}
  \begin{split}
    \mathcal{P}\int_{-k_{\mathrm{min}}}^{k_{\mathrm{max}}}\frac{dk_\alpha P(k_{\alpha})}{\left[k_\alpha + k_{\mathrm{min}} + \delta k_r(\omega) - dA_d\omega^2\right]\left(k_\alpha + k_{\mathrm{min}}\right)}
    &= \frac{d-1}{d[k_{\mathrm{min}}+\delta k_r(\omega)]}\frac{- \frac{\delta k_r(\omega)}{k_{\mathrm{min}}} + \frac{A_d\omega^2}{k_{\mathrm{min}}+\delta k_r(\omega)}}{-\delta k_r(\omega) + dA_d\omega^2}. \\
  \end{split}
\end{equation}
\end{widetext}
where we decompose the real part of the solution as $k_r(\omega) = \left[k_{\mathrm{min}}+\delta k_r(\omega)\right]/(d-1)$.
Because $\delta k_r(\omega)<0$ is expected for consistency with the results in the preceding sections, $-\delta k_r(\omega) + dA_d\omega^2 \neq 0$ for $\omega\neq0$.
Therefore, if the distribution behaves as $P(k_\alpha) \sim (k_\alpha + k_{\mathrm{min}})^\nu$ with $\nu\leq1$ at $k_\alpha \simeq -k_{\mathrm{min}}$, the left-hand side diverges as $\omega$ moves toward zero frequency, whereas the right-hand side is always finite regardless of the frequency dependence of $\delta k_r(\omega)$.
Hence, when the distribution decays with a power $\nu\leq1$ near its lower cutoff, it avoids the local instability.
Finally, note that a distribution with $\nu > 1$ is needed to cause local instability.
This condition is used in Section~\ref{sec:The lowest-frequency region}.

\begin{figure}
    \centering
    \includegraphics[width=0.45\textwidth]{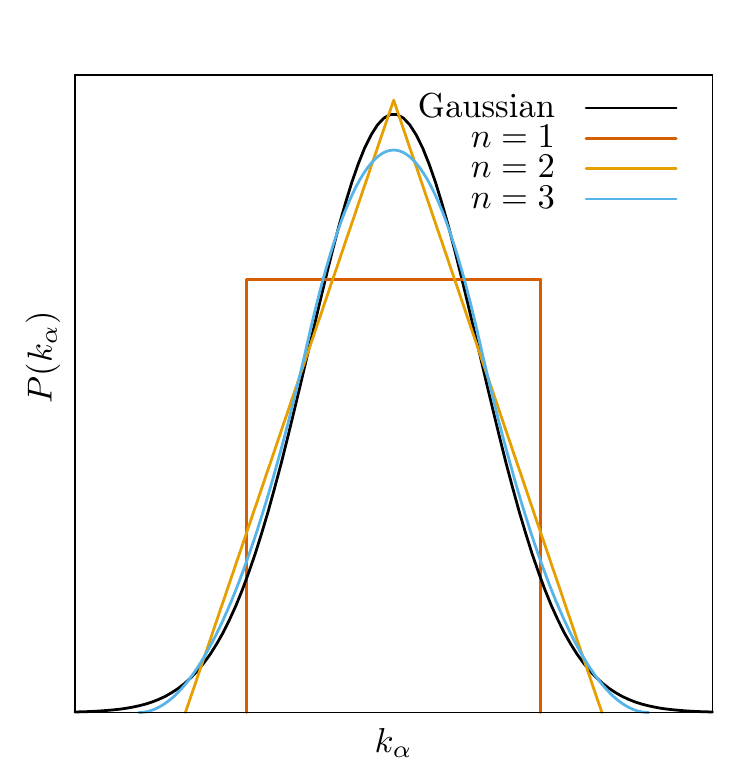}
    \caption{Gaussian and Bates distributions ($n=1, 2$, and $3$) with the same mean and standard deviation.}
    \label{fig:fig3}
\end{figure}

\begin{figure*}[t]
  \begin{center}
    \includegraphics[width=0.9\textwidth]{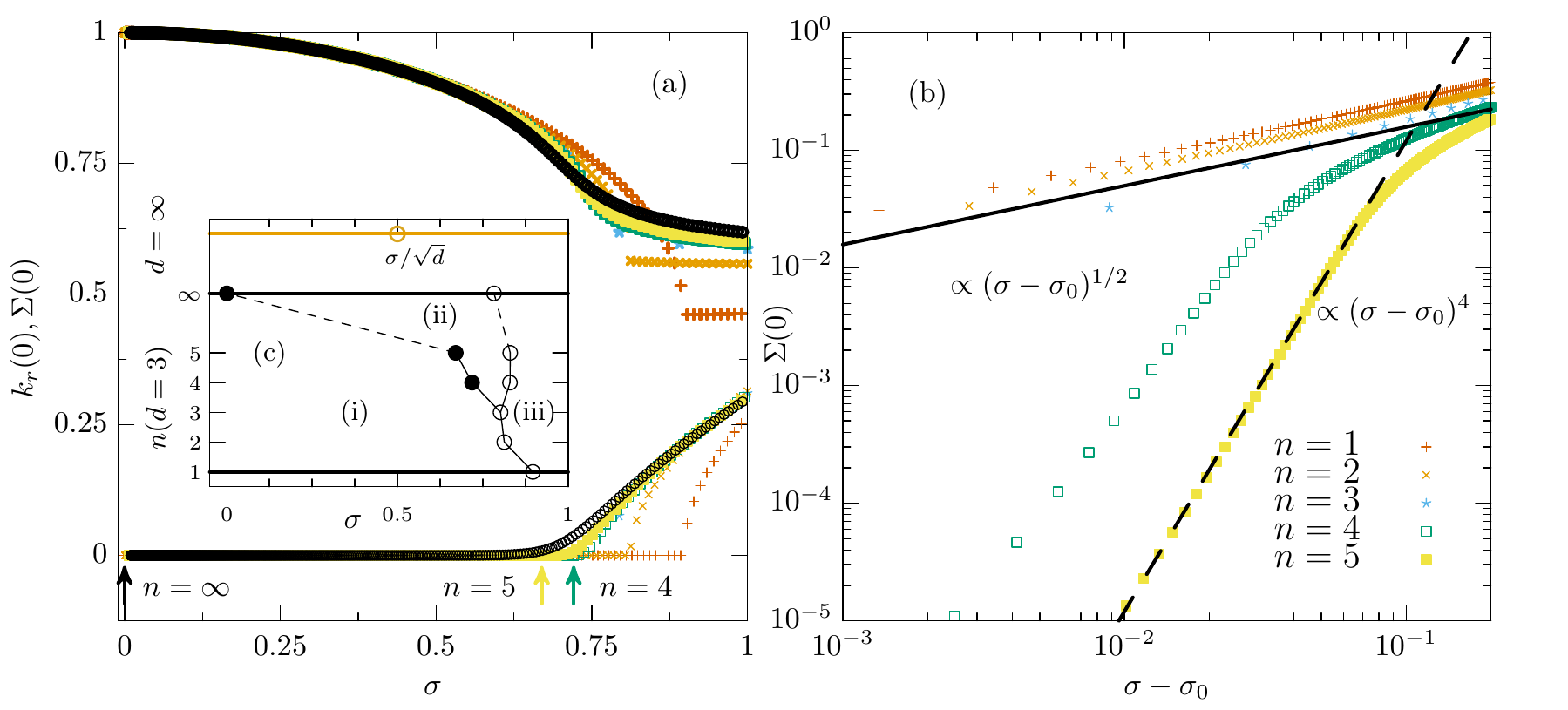}
    \vspace*{0 mm}
    \caption{
      (a) Real and imaginary parts of effective stiffness for the Bates distribution as functions of $\sigma$ at zero frequency.
      The arrows for $n=4,\ 5$, and $\infty$ indicate the critical values above which the system is destabilized by local instability.
      The system with $n\leq3$ is destabilized by conventional instability.
      (b) Logarithmic plot of imaginary parts near the critical values $\sigma=\sigma_0$.
      The solid line indicates $\Sigma(0)\sim(\sigma-\sigma_{0})^{1/2}$.
      The dashed line indicates $\Sigma(0)\sim\left(\sigma-\sigma_{0}\right)^4$.
      (c) Phase diagram for the $(n,\sigma)$ plane in $d=3$.
      The open circles indicate $\sigma_{\mathrm{fit}}$ (see text for definition), and the closed circles indicate values above which local instability occurs.
      The solid and dotted lines indicate visual guidance.
      The plane is divided into three phases: (i) the stable phase, (ii) the unstable phase owing to local instability, and (iii) the unstable phase owing to conventional instability.
      Above the phase diagram in $d=3$, we indicate the phase diagram in $d=\infty$ using the solid line.
      The line is divided into two phases: (i) the stable phase for $\sigma/\sqrt{d}<\sigma_c/\sqrt{d}=\mu/2$ and (ii) the unstable phase owing to conventional instability for $\sigma/\sqrt{d}>\sigma_c/\sqrt{d}$.
    }
    \label{fig:fig4}
  \end{center}
\end{figure*}

\subsection{Bates distribution of stiffness}\label{sec:Bates distribution at zero frequency}

In the preceding Sections~\ref{sec:Gaussian distribution} and~\ref{sec:Restrictions on the probability distribution}, it has been concluded that the Gaussian distribution always induces local instability and is not suitable for the stiffness distribution of a stable system.
This fact is apparent once we observe that the Gaussian distribution has a finite probability of arbitrarily large negative values.
Even if the distribution of the stable system resembles a Gaussian distribution near the mean value, its tail does not infinitely continue and should be cut off at a finite value.

A significant question is which instability generally occurs depending on the stiffness distribution.
Therefore, we employed the Bates distribution, which is a distribution of the average of $n$ statistically independent uniformly distributed random variables in the interval $[\mu-\Delta, \mu+\Delta]$.
The mean is $\mu > 0$, and the variance is $\sigma^2 = \Delta^2/3n$.
It includes the uniform ($n=1$), triangular ($n=2$), and Gaussian ($n,\Delta\to\infty$ and $\sigma = \mathrm{const.}$) distributions.
See Fig.~\ref{fig:fig3}.
Because the Gaussian (uniform) distribution always causes local (conventional) instability, the question can be rephrased as follows: ``Which instability occurs at the general $n$ and $\Delta$?''

To determine the stability of the model, it suffices to calculate the effective stiffness at zero frequency $k_{\mathrm{eff}}(0)$ by solving Eq.~(\ref{eq:zero frequency self consistent equation}).
Figure~\ref{fig:fig4} presents the numerical results in $d=3$.
Figure~\ref{fig:fig4}(a) presents the zero-frequency effective stiffness as a function of the standard deviation $\sigma$ for $n=1,\ 2,\ 3,\ 4,\ 5,$ and $\infty$.
The behaviors of the real parts $k_r(0)\equiv\Re k_{\mathrm{eff}}(0)$ are qualitatively the same as in the preceding sections for all $n$.
When we increase $\sigma$, the imaginary part $\Sigma(0)\equiv-\Im k_{\mathrm{eff}}(0)$ starts to be nonzero at some point $\sigma=\sigma_0$ for a finite $n$, whereas the model with the Gaussian distribution ($n=\infty$) is always unstable.
Note that as $d\to\infty$, we have $\sigma_0\to\sigma_c\equiv\sqrt{d}\mu/2$ as indicated in Section~\ref{sec:Large dimension limit}.
If the system is destabilized by the local instability, Eq.~(\ref{eq:stability condition}) is violated above $\sigma=\sigma_0$.
We found that the models with $n=4$ and $5$ are destabilized by the local instability; the instability points are indicated by the arrows in the figure.
We also represent the instability point in the Gaussian case, that is, $\sigma_0=0$.

On the other hand, the models with $n=1, 2,$ and $3$ are destabilized without violating Eq.~(\ref{eq:stability condition}).
Because the model with the uniform distribution ($n=1$) is destabilized by the conventional instability even in $d=3$ as shown in Section~\ref{sec:Uniform distribution}, this result indicates that the models with $n=1$ and $2$ are also destabilized by the conventional instability.
Moreover, when $\sigma$ is increased, we expect a crossover from the local instability to the conventional instability even in $n=4$ and $5$ as in the case of the Gaussian distribution (see Section~\ref{sec:Gaussian distribution}).
Actually, in the region of the largest standard deviation in Fig.~\ref{fig:fig4}(a), the imaginary parts for all $n$ almost overlap.

To corroborate this observation, we present the logarithmic plot of the imaginary parts near $\sigma=\sigma_0$ in Fig.~\ref{fig:fig4}(b).
The solid line in the figure indicates $\Sigma(0)\sim\left(\sigma-\sigma_0\right)^{1/2}$, and the dashed line indicates $\Sigma(0)\sim\left(\sigma-\sigma_0\right)^4$.
The former power law is the $\sigma$-dependence in the large-dimension limit shown in Eq.~(\ref{eq:infinite dimensional solution omega_0}).
Thus, the fact that the data for $n\leq3$ are fitted well to $\Sigma(0)\sim\left(\sigma-\sigma_0\right)^{1/2}$ indicates that these models are destabilized by the conventional instability, similar to the large-dimension limit.
For $n\geq4$, by contrast, the $\sigma$-dependence seems to depend on $n$, e.g., $\Sigma(0)\sim\left(\sigma-\sigma_0\right)^4$ for $n=5$.
This is characteristic of the local instability.
Furthermore, we attempted to fit our numerical solutions using the $\sigma$-dependence of the conventional instability, $\Sigma(0) = A\sqrt{\sigma^2 - \sigma_{\mathrm{fit}}^2}$ with two fitting parameters $A$ and $\sigma_{\mathrm{fit}}$ \footnote{We fitted the regions distant from $\sigma_0$ for $n=4, 5$ and $\infty$ to avoid the effects of the local instability.}.
For $n=1,\ 2,$ and $3$, a sufficient fit is obtained, and $\sigma_{\mathrm{fit}}\simeq\sigma_0$.
For a larger $n$, a sufficient fit is obtained only for the large $\sigma$, and $\sigma_{\mathrm{fit}}(>\sigma_0)$ should be interpreted as the estimate of the crossover point from the local instability to the conventional instability.

Finally, Fig.~\ref{fig:fig4}(c) summarizes the critical and crossover points in $d=3$ and $d=\infty$.
For $d=3$, the open symbols indicate $\sigma_{\mathrm{fit}}$, which is determined by fitting, and the closed symbols indicate $\sigma_0$ for $n\geq4$.
These values allow us to draw the phase diagram in the $(n,\sigma)$ plane.
The plane is divided into the following three phases: (i) the stable phase, (ii) unstable phase owing to local instability, and (iii) unstable phase owing to conventional instability.
On the other hand, for $d=\infty$, the phase diagram is simple: (i) the stable phase at $\sigma/\sqrt{d}<\sigma_c/\sqrt{d}=\mu/2$ and (ii) the unstable phase owing to the conventional instability at $\sigma/\sqrt{d}>\sigma_c/\sqrt{d}$.

\subsection{Marginal solution by local instability}\label{sec:The lowest-frequency region}

In the preceding section, we found that the distributions supported on the finite interval $[-k_{\mathrm{min}}, k_{\mathrm{max}}]$ can exhibit local instability.
A next issue is the frequency dependence of a solution when the system is marginally stable owing to local instability.

To address this issue, we focus on the lowest-frequency region at the critical standard deviation, that is, $k_{\mathrm{eff}}(0) = k_{\mathrm{min}}/(d-1)$. We decompose the real part of the solution $k_r(\omega) \equiv \Re k_{\mathrm{eff}}(\omega) = \left[k_{\mathrm{min}} + \delta k_r(\omega)\right]/(d-1)$ (see Section~\ref{sec:Restrictions on the probability distribution}).
We also approximate Green's function using Eq.~(\ref{eq:approximate Green's function})~\cite{Degiuli2014Force}.
Thus, the equation for $k_r(\omega)$ is Eq.~(\ref{eq:equation for the real part}).
After solving Eq.~(\ref{eq:equation for the real part}), the imaginary part $\Sigma(\omega) \equiv -\Im k_{\mathrm{eff}}(\omega)$ is given by
\begin{equation}\label{eq:equation for the imaginary part}
  \Sigma(\omega) = \pi dk_r(\omega)^2P\left[-(d-1)k_r(\omega) + dA_d\omega^2\right].
\end{equation}

We solve Eq.~(\ref{eq:equation for the real part}) under the conditions $k_{\mathrm{eff}}(0) = k_{\mathrm{min}}/(d-1)$ and $P(k_\alpha) \sim (k_\alpha + k_{\mathrm{min}})^\nu$ with $\nu>1$ at $k_\alpha\sim -k_{\mathrm{min}}$ in Appendix~\ref{sec:Marginal solution by the local instability}.
The result is 
\begin{equation}\label{eq:delta k}
  \frac{\delta k_r(\omega)}{d-1} = -\frac{dI_2 - dI_1^2}{dI_1^2 - (d-1)I_2}A_d\omega^2.
\end{equation}
with
\begin{equation}
  I_{m} \equiv \int_{-1}^1dx\frac{\Delta P(\Delta x + \mu)}{(x+1)^m},
\end{equation}
where $k_{\mathrm{min}} = \Delta - \mu$ and  $k_{\mathrm{max}} = \Delta + \mu$, as in Sections~\ref{sec:Uniform distribution} and~\ref{sec:Bates distribution at zero frequency}.
We note, however, that our argument in this section is not restricted to the Bates distribution.
Because its numerator is always non-negative, an inequality
\begin{equation}\label{eq:inequality}
  d < \frac{I_2}{I_2 - I_1^2} \equiv d_t,
\end{equation}
holds for $\delta k_r(\omega)$ to be negative, which is required for the local minimum of the sound velocity~\cite{Wyart2010Scaling, Degiuli2014Effects}.

Finally, when the real part of the solution $\delta k_r(\omega)$ is substituted into Eq.~(\ref{eq:equation for the imaginary part}), we obtain
\begin{equation}
  \Sigma(\omega) \sim \omega^{2\nu}.
\end{equation}
Its contribution to the vDOS is
\begin{equation}\label{eq:general law}
  g_{\mathrm{local}}(\omega) \sim \omega^{2\nu+1},
\end{equation}
which is clearly different from the Debye behavior $g_{\mathrm{Debye}}(\omega)\sim\omega^{d-1}$.
In the preceding study~\cite{Shimada2019Vibrational}, we presented a model that shows the quartic law $g_{\mathrm{local}}(\omega)\sim\omega^4$, which is a special case of $\nu=3/2$ in Eq.~(\ref{eq:general law}).

\section{Vector displacement model}~\label{Vector displacement model}

\subsection{Model}\label{sec:Model2}

Until now, we have focused on the SDM.
It is analytically simple and is sufficient to qualitatively analyze the vibrations of glass.
However, a slightly realistic model with the vector displacements $\{\boldsymbol{u}_i\}_{i=1}^N$ can be treated.
We refer to this as the VDM, to differentiate it from the SDM.
Because the VDM has been widely used~\cite{Wyart2010Scaling,Degiuli2014Effects,Degiuli2014Force}, its results are presented here for convenience.

The equation of motion is
\begin{equation}
    \frac{d^2}{dt^2}\boldsymbol{u}_i = -\sum_{j\in\partial i}k_{ij}\boldsymbol{n}_{ij}\boldsymbol{n}_{ij}\cdot(\boldsymbol{u}_{i}-\boldsymbol{u}_j),
\end{equation}
where $\boldsymbol{n}_{ij}$ is the unit vector from the $i$th element to the $j$th element.
In the bra-ket notation,
\begin{equation}
    \frac{d^{2}}{d t^{2}}\ket{u}=-\hat{\mathcal{M}}\ket{u},
\end{equation}
where
\begin{equation}
    \begin{split}
        \hat{\mathcal{M}} &= \sum_{\left<ij\right>}k_{ij}(\ket{i}-\ket{j})\boldsymbol{n}_{ij}\boldsymbol{n}_{ij}^T(\bra{i}-\bra{j})\\
        &\equiv \sum_{\alpha=\left<ij\right>}k_\alpha\ket{\alpha}\boldsymbol{n}_\alpha\boldsymbol{n}_\alpha^T\bra{\alpha}.
    \end{split}
\end{equation}
Note that a simple cubic lattice cannot be chosen for the VDM because it does not have a finite shear modulus.
Specifically, a lattice coordination number $z$ must be greater than the Maxwell criterion $2d$, below which the network loses rigidity.

Green's function for the VDM differs by the choice of the lattice; however, low-frequency properties are expected to be universal and do not depend on the lattice.
Thus, we consider a simplified Green's function for a homogeneous system
\begin{equation}
    \begin{split}
        \hat{G}_K(\boldsymbol{r}_{ij},\omega) &\equiv \bra{i}\hat{\mathcal{G}}_K(\omega)\ket{j} \\ 
        &= \int_0^{q_D}\frac{d\boldsymbol{q}}{(2\pi)^d}\frac{e^{i\boldsymbol{q}\cdot\boldsymbol{r}_{ij}}}{K\boldsymbol{q}^2-\omega^2}\hat{\delta}_d,
    \end{split}
\end{equation}
where $\hat{\delta}_d$ is the $d\times d$ identity matrix~\cite{Wyart2010Scaling,During2013Phonon,Degiuli2014Effects,Degiuli2014Force}.
This is the same as Green's function for the SDM except for the factor $\hat{\delta}_d$.

Similar to the SDM, we obtain the self-consistent equation for the VDM
\begin{equation}\label{eq:self consistent equation for the VDM}
    \overline{\frac{k_{\mathrm{eff}}(\omega)-k_\alpha}{1 - \left[k_{\mathrm{eff}}(\omega)-k_\alpha\right]\frac{\theta}{k_{\mathrm{eff}}(\omega)}\left[1 + \omega^2G(\omega)\right]}} = 0,
\end{equation}
where
\begin{equation}
    G(\omega) = \frac{1}{d}\Tr \hat{G}_{\mathrm{eff}}(\boldsymbol{0},\omega) = \int_0^{q_D}\frac{d\boldsymbol{q}}{(2\pi)^d}\frac{1}{k_{\mathrm{eff}}(\omega)\boldsymbol{q}^2-\omega^2},
\end{equation}
and $\theta = 2n_{\mathrm{dof}}/z$.
$n_{\mathrm{dof}}$ is the number of degrees of freedom per element and $n_{\mathrm{dof}} = d$ in the VDM.
For the SDM in the simple cubic lattice, $z=2d$ and $n_{\mathrm{dof}} = 1$; thus, $\theta = 1/d$.
Essentially, when $\theta$ is replaced by $1/d$, we reproduce Eq.~(\ref{eq:self consistent equation}).

\subsection{Difference from SDM}\label{sec:Difference from the scalar displacement model}

The self-consistent equations for the VDM and SDM are almost the same.
In this section, we present how the results for the SDM are modified for the VDM.
In the zero-frequency limit, we can obtain the results for the VDM by replacing $d$ in the results of the SDM by $1/\theta$.
For example, Eq.~(\ref{eq:stability condition}) is modified as follows: 
\begin{equation}
    P\left[-(\theta-1)k_{\mathrm{eff}}(0)/\theta\right] = 0,
\end{equation}
which was already reported in the preceding study~\cite{Shimada2019Vibrational}.
In the case of finite frequency, $d$ needs to be replaced with $1/\theta$, except for $A_d$ defined in Eq.~(\ref{eq:approximate Green's function}).
Therefore, Eq.~(\ref{eq:delta k}) is modified as follows: 
\begin{equation}
    \frac{\delta k_r(\omega)}{\theta^{-1}-1} = -\frac{I_2 - I_1^2}{I_1^2 - (1-\theta)I_2/\theta}A_d\omega^2,
\end{equation}
Thus, Eq.~(\ref{eq:inequality}) becomes a condition for $\theta$: 
\begin{equation}
    \theta > 1 - \frac{I_1^2}{I_2} \equiv \theta_t.
\end{equation}

Moreover, the large-dimension limit of the VDM needs to be discussed.
For the SDM, the self-consistent equation can be expanded, as shown in Section~\ref{sec:Large dimension limit}, in the large-dimension limit.
To do the same for the VDM, $\theta\ll1$ is needed rather than $d\gg1$.
Therefore, the large-dimension limits of the SDM and VDM do not necessarily correspond.

Finally, we note an additional benefit of the VDM.
In this model, the parameter $\theta$ can be changed independent of the spatial dimension $d$.
This corresponds to changing the lattice.
Thus, $\theta$ can be considered a control parameter.
This enables us to present an anomalous model in which the effective stiffness vanishes at zero frequency: $k_{\mathrm{eff}}(0)=0$ when we appropriately choose the value of $\theta$.
In Appendix~\ref{sec:Model with the continuously vanishing stiffness}, we provide this type of model and present its relation to nearly jammed materials.

\section{Summary and discussion}\label{sec:Summary and discussion}

In this study, we extended the analysis conducted in the preceding study~\cite{Shimada2019Vibrational} and derived new results for local instability.
In the first part of this paper, we analyzed the SDM, which is one of the simplest elasticity models, using the EMA.
We first considered the large-dimension limit, where the EMA becomes exact~\cite{Luck1991Conductivity}, and determined that the model yields the gapless non-Debye scaling law, $g(\omega)\sim\omega^2$, when the system is marginally stable.
Therefore, the non-Debye scaling and the associated conventional instability originate from the purely mean-field nature.
Next, we analyzed the SDM with specific distributions of stiffness, a uniform distribution, and a Gaussian distribution.
The uniform distribution yields approximately the same results as in the large-dimension limit in relatively small dimensions, whereas the Gaussian distribution always leads to an unstable solution.

Considering the difference between the uniform and Gaussian distributions, a local instability was introduced, and its relationship with the QLVs was analyzed.
In particular, we observed similarities in the size and energetics between the cores of the QLVs and the response to a local dipolar force.
In real amorphous solids, the size of these cores is larger than the microscopic particle size, whereas it suffices to consider one spring in the proposed elasticity models.
Therefore, the appropriate coarse-graining length for the elasticity theory of amorphous solids is expected to be the size of the QLVs~\cite{Lerner2014Breakdown}.
That is, once the atomistic system is coarse-grained to the size of the QLVs, it reduces to the elasticity model.
In contrast, for scales below this length, the microscopic motions of the constituent particles need to be considered.

Note that there may be other types of instabilities in addition to the local and conventional instabilities under the EMA.
We cannot prove that conventional instability always occurs when avoiding local instability.
However, as indicated in Sections~\ref{sec:Large dimension limit} and~\ref{sec:Uniform distribution}, the imaginary part arises as the solution of the quadratic self-consistent equation in the case of conventional instability, which generally occurs if we can expand the equation in a series.
When we go beyond the EMA, the combinations of transfer matrices yield several other types of instabilities.
However, numerical or experimental studies have not detected such complicated instabilities in real glasses.
Therefore, we expect that the elasticity theory with quenched disorder under the EMA is a suitable framework for describing amorphous solids and that higher-order terms in Eq.~(\ref{eq:T matrix series}) do not significantly improve our understanding.

Having introduced the local instability, we illustrated that the tail of the stiffness distribution needs to decay more rapidly than linearly with $(k+k_{\mathrm{min}})$, to cause local instability.
Based on this result, we considered the Bates distribution, which covers a wide range of distributions, and illustrated that the SDM with the Bates distribution is destabilized by the local instability as well as the conventional instability, depending on the specific shape of the distribution.
Concretely, we found that the local (conventional) instability tends to occur when the distribution resembles a Gaussian (uniform) distribution.
Thus, we can determine the ``phase diagram'' in the parameter space of the distribution in which the stable phase and the unstable phases by the local and conventional instabilities are separated.
We also illustrated that when the system is on the verge of local instability, the vDOS follows another power law: $g_{\mathrm{local}}(\omega)\sim\omega^{2\nu+1}$, where $\nu>1$ is the exponent of the distribution.
This is consistent with the exponent of the vDOS of the QLVs $\beta = 4$~\cite{Lerner2017Effect, Lerner2020Finite-size}, as indicated in the Introduction.

In the second part of the study, we presented the VDM.
Here, note that in Refs.~\cite{Degiuli2014Effects,Degiuli2014Force}, the initial stress was considered as the source of the instability.
Specifically, the term $-\frac{f_{\alpha}}{\left|\boldsymbol{r}_{\alpha}\right|}\ket{\alpha}\left(\hat{\delta}_{d}-\boldsymbol{n}_{\alpha} \otimes \boldsymbol{n}_{\alpha}\right)\bra{\alpha}$, where $f_\alpha$ is the force between a pair $\alpha$, was considered in the dynamical matrix.
Although the initial stress was neglected in our study, the mechanism of the local instability is general and valid even when the initial stress is considered
In this case, the distribution of $f_\alpha$ plays a central role.

Throughout the present analyses regarding the two types of elasticity models, our main finding is that local instability can occur in finite spatial dimensions.
The unstable modes induced by the local instability share characteristic features with the QLV modes, lying in the low-frequency edge of the spectrum.
Moreover, while the conventional instability robustly produces the non-Debye quadratic law of $g(\omega) \propto \omega^2$, the quartic law of $g(\omega) \propto \omega^4$, which has been observed in many finite-dimensional systems~\cite{Shimada2018Anomalous,Lerner2016Statistics,Mizuno2017Continuum,Wang2019Lowfrequency,Richard2020Universality,Prasenjit2020Robustness}, can be rationalized in terms of local instability.
In addition to this vDOS power law, because the distribution of local elastic moduli appears to be significantly similar to a Gaussian distribution~\cite{Mizuno2013Measuring}, real amorphous solids are expected to be governed by the local instability based on the analysis of the phase diagram in Fig.~\ref{fig:fig4}(c).
This argument reinforces the fact that the local instability yields low-frequency modes which share the same properties with the QLVs.
Therefore, we propose that finite-dimensional amorphous solids are in the marginally stable phase in terms of local instability.
However, we cannot prove that $\nu$ must be $3/2$ or $2\nu+1=4$, within the elasticity theory.
This is natural because the precise functional form of $P(k_\alpha)$ should be determined by the dynamics in which the system freezes into a solid state.
Investigating this time evolution of $P(k_\alpha)$ upon cooling may be considered for a future study.

The local nature of marginal stability is also consistent with numerical observations in that amorphous solids undergo \textit{local} rearrangements under mechanical loading or thermal agitation~\cite{Karmakar2010Statistical,Mizuno2020anharmonic} that involve $10$ to $1000$ particles.
Experimentally, it is relatively difficult to directly observe these microscopic phenomena in samples of materials.
However, they can be detected indirectly through experimental measurements.
For example, the anomalous temperature dependence of the heat capacity and the thermal conductivity~\cite{Zeller1971Thermal,Phillips1981Amorphous,Perez-Castaneda2_2014} is considered to be caused by localized transitions through the quantum tunneling mechanism, i.e., the two-level systems~\cite{Perez-Castaneda2_2014,Anderson1972Anomalous,Phillips1972Tunneling}.

In future studies, it will be interesting to investigate the effects of anharmonicity or the correlation of the local elastic modulus.
The former is necessary to directly analyze the yielding transition~\cite{Maloney2006,Tanguy2010,Manning2011,Karmakar2010Statistical} and thermal properties of glasses~\cite{Zeller1971Thermal,Anderson1972Anomalous,Phillips1972Tunneling,Phillips1981Amorphous,Karpov1983Theory,Buchenau1991Anharmonic,Buchenau1992Interaction}.
Recently, the phenomenological theory has been proposed to derive the vDOS of the QLVs by treating the anharmonicity~\cite{Ji2019Theory}. 
The latter has been analyzed for several decades and has implications for phonon transport~\cite{John1983Wave,Cui2019Analytical}; its understanding has been advanced by recent numerical simulations~\cite{Mizuno2018phonon,Wang2019sound,Moriel2019wave}.

\section*{Acknowledgments}

This study was supported by JSPS KAKENHI Grant Numbers 19J20036, 17H04853, 18H05225, 18H03675, 19H01812, 19K14670, 20H01868, and 20H00128.
It was also partially supported by the Asahi Glass Foundation.

\onecolumngrid
\appendix

\section{Approximation of Green's function in large-dimension limit}\label{sec:Approximation of the Green's function in the large dimension limit}
In this appendix, Eq.~(\ref{eq:approximate Green's function}) is derived using the Debye approximation.
Introducing polar coordinates, $G\left(\omega\right)$ is expressed as follows:
\begin{equation}
  \begin{split}
    G\left(\omega\right) &= \int_{0<|\boldsymbol{q}|<q_D}\frac{d\boldsymbol{q}}{\left(2\pi\right)^d}\frac{1}{k_{\mathrm{eff}}(\omega)\boldsymbol{q}^2 - \omega^2} \\
    &= \frac{S_{d-1}}{\left(2\pi\right)^d}\int_0^{q_D}dq\frac{q^{d-1}}{k_{\mathrm{eff}}(\omega)q^2 - \omega^2} \\
    &= \frac{S_{d-1}q_D^d}{\left(2\pi\right)^d}\int_0^1dq\frac{q^{d-1}}{k_{\mathrm{eff}}(\omega)q_D^2q^2 - \omega^2} \\
    &= \frac{d}{k_{\mathrm{eff}}(\omega)q_D^2}\int_0^1dq\frac{q^{d-1}}{q^2 - \omega^2/k_{\mathrm{eff}}(\omega)q_D^2}, \\
  \end{split}
\end{equation}
We have used Eq.~(\ref{eq:Debye wavenumber}) in the last line.
Because an expansion in powers of frequency is convenient, the integrand is expanded as follows:
\begin{equation}
  \frac{q^{d-1}}{q^2 - \omega^2/k_{\mathrm{eff}}(\omega)q_D^2} =
  \begin{cases}
    \sum_{n=0}^{D-1} q^{2D-2-2n} \left[\frac{\omega^2}{k_{\mathrm{eff}}(\omega)q_D^2}\right]^{n} + \frac{1}{q^{2}-{\omega^2}/{k_{\mathrm{eff}}(\omega)q_D^2}}\left[\frac{\omega^2}{k_{\mathrm{eff}}(\omega)q_D^2}\right]^{D} & (d = 2D+1,D=1,2,\cdots) \\
    \sum_{n=0}^{D-1} q^{2D-1-2n} \left[\frac{\omega^2}{k_{\mathrm{eff}}(\omega)q_D^2}\right]^{n} + \frac{q}{q^{2}-{\omega^2}/{k_{\mathrm{eff}}(\omega)q_D^2}}\left[\frac{\omega^2}{k_{\mathrm{eff}}(\omega)q_D^2}\right]^{D} & (d = 2D+2,D=1,2,\cdots) \\
  \end{cases}
  .
\end{equation}
As $d\to\infty$, the second terms can be neglected in both cases. 
Thus, we obtain
\begin{equation}
  \int_0^1dq\frac{q^{d-1}}{q^2 - \omega^2/k_{\mathrm{eff}}(\omega)q_D^2} \simeq
  \begin{cases}
    \sum_{n=0}^{D-1} \frac{1}{d-2-2n} \left[\frac{\omega^2}{k_{\mathrm{eff}}(\omega)q_D^2}\right]^{n} & (d = 2D+1,D=1,2,\cdots) \\
    \sum_{n=0}^{D-1} \frac{1}{d-2-2n} \left[\frac{\omega^2}{k_{\mathrm{eff}}(\omega)q_D^2}\right]^{n} & (d = 2D+2,D=1,2,\cdots) \\
  \end{cases}
  .
\end{equation}
Because only the expression that is valid in the low-frequency region is needed, we approximate $1/(d-2-2n) \sim 1/d$, and therefore obtain
\begin{equation}
  G\left(\omega\right) = \frac{d}{k_{\mathrm{eff}}(\omega)q_D^2}\int_0^1dq\frac{q^{d-1}}{q^2 - \omega^2/k_{\mathrm{eff}}(\omega)q_D^2} \simeq \frac{1}{k_{\mathrm{eff}}(\omega)q_D^2}\sum_{n=0}^{\infty} \left[\frac{\omega^2}{k_{\mathrm{eff}}(\omega)q_D^2}\right]^{n} = \frac{1}{k_{\mathrm{eff}}(\omega)q_D^2 - \omega^2}.
\end{equation}

\section{Solution for uniform distribution}\label{sec:Solution for the uniform distribution}

In this appendix, we solve Eq.~(\ref{eq:self consistent equation for uniform distribution}).
To solve the cubic equation, we substitute $k_{\mathrm{eff}}(\omega) = y + \mu/3$,
\begin{equation}
  \begin{split}
    & k_{\mathrm{eff}}(\omega)^3 -\mu k_{\mathrm{eff}}(\omega)^2 + \frac{\sigma^2}{d}k_{\mathrm{eff}}(\omega) + \frac{\sigma^2}{d}A_d\omega^2 \\
    &=\left(y + \frac{\mu}{3}\right)^3 -\mu \left(y + \frac{\mu}{3}\right)^2 + \frac{\sigma^2}{d}\left(y + \frac{\mu}{3}\right) + \frac{\sigma^2}{d}A_d\omega^2 \\
    &= y^3 + \left(\frac{\mu^2}{3} - \frac{2\mu^2}{3} + \frac{\sigma^2}{d}\right)y + \frac{\mu^3}{27} - \frac{\mu^3}{9} + \mu\frac{\sigma^2}{3d} + \frac{\sigma^2}{d}A_d\omega^2 \\
    &= y^3 + \left(-\frac{\mu^2}{3} + \frac{\sigma^2}{d}\right)y - \frac{2\mu^3}{27} + \mu\frac{\sigma^2}{3d} + \frac{\sigma^2}{d}A_d\omega^2 \\
    &= y^3 + 3\left(-\frac{\mu^2}{9} + \frac{\sigma^2}{3d}\right)y  + 2\left(- \frac{\mu^3}{27} + \mu\frac{\sigma^2}{6d} + \frac{\sigma^2}{2d}A_d\omega^2\right)\\
    &\equiv y^3 + 3Py + 2Q.
  \end{split}
\end{equation}
Using the critical value defined in Section~\ref{sec:Large dimension limit}, $\sigma_c = \sqrt{d}\mu/2$, $P$ can be written as follows:
\begin{equation}
    \begin{split}
    P & = -\frac{\mu^2}{9} + \frac{\sigma^2}{3d} \\
    &= -\frac{\mu^2}{9}  + \frac{\sigma_c^2}{3d} - \frac{\sigma_c^2-\sigma^2}{3d} \\
    &= -\frac{\mu^2}{36} - \frac{\sigma_c^2-\sigma^2}{3d} \\
    &= -\left(\frac{\mu}{6}\right)^2 - \frac{\sigma_c^2-\sigma^2}{3d} \\
    &\equiv -\left(\frac{\mu}{6}\right)^2 - \frac{\delta_\sigma^2}{3d}. \\
  \end{split}
\end{equation}
Likewise, $Q$ becomes
\begin{equation}
  \begin{split}
    Q &= -\frac{\mu^3}{27} + \mu\frac{\sigma^2}{6d} + \frac{\sigma^2}{2d}A_d\omega^2  \\
    &= -\frac{\mu^3}{27} + \mu\frac{\sigma_c^2}{6d} - \mu\frac{\sigma_c^2 - \sigma^2}{6d} + \frac{\sigma^2}{2d}A_d\omega^2 \\
    &= \left(-\frac{1}{27}+\frac{1}{24}\right)\mu^3 - \mu\frac{\sigma_c^2 - \sigma^2}{6d} + \frac{\sigma^2}{2d}A_d\omega^2  \\
    &= \left(\frac{\mu}{6}\right)^3 - \mu\frac{\sigma_c^2 - \sigma^2}{6d} + \frac{\sigma^2}{2d}A_d\omega^2  \\
    &\equiv \left(\frac{\mu}{6}\right)^3 - \frac{\mu\delta_\sigma^2}{6d} + \frac{\sigma^2}{2d}A_d\omega^2.  \\
  \end{split} 
\end{equation}
 To analyze the vibrations for $\sigma\lesssim\sigma_c$, we set $\delta_\sigma^2/\mu^2 \sim \omega^2/\mu \ll 1$ and derive an approximate expression for $k_{\mathrm{eff}}(\omega)$.
Therefore, $Q$ can be approximated as follows:
\begin{equation}
  Q \simeq \left(\frac{\mu}{6}\right)^3 - \frac{\mu\delta_\sigma^2}{6d} + \frac{\sigma_c^2}{2d}A_d\omega^2.
\end{equation}
Next, we need to compute
\begin{equation}
  \left(-Q \pm \sqrt{Q^2 + P^3}\right)^{1/3}.
\end{equation}
The terms under the square root are 
\begin{equation}
  \begin{split}
    & Q^2 + P^3 \\
    &\simeq \left[\left(\frac{\mu}{6}\right)^3 - \frac{\mu\delta_\sigma^2}{6d} + \frac{\sigma_c^2}{2d}A_d\omega^2\right]^2 - \left[\left(\frac{\mu}{6}\right)^2 + \frac{\delta_\sigma^2}{3d}\right]^3 \\
    &= \left(\frac{\mu}{6}\right)^6 - 2\left(\frac{\mu}{6}\right)^3\frac{\mu\delta_\sigma^2}{6d} + \left(\frac{\mu\delta_\sigma^2}{6d}\right)^2 + \frac{\sigma_c^2}{2d}A_d\omega^2\left\{\frac{\sigma_c^2}{2d}A_d\omega^2 + 2\left[\left(\frac{\mu}{6}\right)^3 - \mu\frac{\delta_\sigma^2}{6d}\right]\right\} \\
    &-\left[\left(\frac{\mu}{6}\right)^6 + 3\left(\frac{\mu}{6}\right)^4\frac{\delta_\sigma^2}{3d} + 3\left(\frac{\mu}{6}\right)^2\left(\frac{\delta_\sigma^2}{3d}\right)^2 + \left(\frac{\delta_\sigma^2}{3d}\right)^3\right] \\
    &=  - 3\left(\frac{\mu}{6}\right)^4\frac{\delta_\sigma^2}{d} + \frac{2}{3}\left(\frac{\mu\delta_\sigma^2}{6d}\right)^2 + \frac{1}{4}\left(\frac{\sigma_c^2}{d}A_d\omega^2\right)^2 + \left(\frac{\mu}{6}\right)^3\frac{\sigma_c^2}{d}A_d\omega^2 - \frac{\mu\delta_\sigma^2}{6d}\frac{\sigma_c^2}{d}A_d\omega^2 \\
    &\simeq  - 3\left(\frac{\mu}{6}\right)^4\frac{\delta_\sigma^2}{d} + \left(\frac{\mu}{6}\right)^3\frac{\sigma_c^2}{d}A_d\omega^2  \\
    &= - 3\left(\frac{\mu}{6}\right)^4\frac{\delta_\sigma^2}{d} + \left(\frac{\mu}{6}\right)^3\frac{\mu^2}{4}A_d\omega^2  \\
    &= 9\left(\frac{\mu}{6}\right)^5\left(A_d\omega^2 - 2\frac{\delta_\sigma^2}{d\mu}\right).  \\
  \end{split}
\end{equation}
Therefore, we obtain 
\begin{equation}
  \begin{split}
    &\left(-Q \pm \sqrt{Q^2 + P^3}\right)^{1/3}\\
    &\simeq \left[\left(-\frac{\mu}{6}\right)^3 + \frac{\mu\delta_\sigma^2}{6d} - \frac{\sigma_c^2}{2d}A_d\omega^2 \pm \sqrt{9\left(\frac{\mu}{6}\right)^5\left(A_d\omega^2 - 2\frac{\delta_\sigma^2}{d\mu}\right)}\right]^{1/3} \\
    &\simeq \left[\left(-\frac{\mu}{6}\right)^3 \pm \sqrt{9\left(\frac{\mu}{6}\right)^5\left(A_d\omega^2 - 2\frac{\delta_\sigma^2}{d\mu}\right)}\right]^{1/3} \\
    &= -\frac{\mu}{6}\left[1 \pm 3\frac{-6}{\mu}\sqrt{\frac{\mu}{6}\left(A_d\omega^2 - 2\frac{\delta_\sigma^2}{d\mu}\right)}\right]^{1/3} \\
    &\simeq -\frac{\mu}{6}\left[1 \pm \frac{-6}{\mu}\sqrt{\frac{\mu}{6}\left(A_d\omega^2 - 2\frac{\delta_\sigma^2}{d\mu}\right)}\right] \\
    &= -\frac{\mu}{6} \pm \sqrt{\frac{\mu}{6}\left(A_d\omega^2 - 2\frac{\delta_\sigma^2}{d\mu}\right)}.
  \end{split}
\end{equation}
A solution that satisfies $k_{\mathrm{eff}}(0)\to\mu$ and $\Sigma(\omega) \equiv -\Im k_{\mathrm{eff}}(\omega) < 0$ as $\sigma\to0$ is chosen.
Thus, we have
\begin{equation}
  \begin{split}
    k_{\mathrm{eff}}(\omega) &= \frac{\mu}{3} + \frac{-1-\sqrt{3}i}{2}\left(-Q+\sqrt{Q^2 + P^3}\right)^{1/3} + \frac{-1+\sqrt{3}i}{2}\left(-Q-\sqrt{Q^2 + P^3}\right)^{1/3} \\
    &\simeq \frac{\mu}{3} + \frac{-1-\sqrt{3}i}{2}\left[-\frac{\mu}{6} + \sqrt{\frac{\mu}{6}\left(A_d\omega^2 - 2\frac{\delta_\sigma^2}{d\mu}\right)}\right] + \frac{-1+\sqrt{3}i}{2}\left[ -\frac{\mu}{6} - \sqrt{\frac{\mu}{6}\left(A_d\omega^2 - 2\frac{\delta_\sigma^2}{d\mu}\right)}\right] \\
    &= \frac{\mu}{2} - i\sqrt{\frac{\mu}{2}\left(A_d\omega^2 - 2\frac{\delta_\sigma^2}{d\mu}\right)} \\
    &\equiv \frac{\mu}{2} - i\sqrt{\frac{\mu}{2}}\sqrt{A_d\omega^2 - A_d{\omega'_{0}}^2}.
  \end{split}
\end{equation}

\section{Asymptotic solution for Gaussian distribution in large-dimension limit}\label{sec:Asymptotic solution for the Gaussian distribution in the large dimension limit}

In this appendix, we derive Eqs.~(\ref{eq:Gaussian real}) and~(\ref{eq:Gaussian imaginary}).
When $\sigma<\sigma_c$ and $d\gg1$, $k_r(0)\gg\Sigma(0)$ can be assumed.
Therefore, linearizing Eq.~(\ref{eq:gaussian self consistent zero frequency}) near $\Sigma(0)$, we obtain
\begin{equation}\label{eq:linearized self consistent equation}
  \begin{split}
    F \left[ \frac { \mu + ( d - 1 ) k_r(0) } { \sqrt { 2 } \sigma } \right] &+ F ^ { \prime } \left[ \frac { \mu + ( d - 1 ) k_r(0) } { \sqrt { 2 } \sigma } \right] \left[ - i \frac { ( d - 1 ) \Sigma(0) } { \sqrt { 2 } \sigma } \right]
    + i \frac { \sqrt { \pi } } { 2 } \exp \left\{ - \frac { \left[ \mu + ( d - 1 ) k_r(0) \right] ^ { 2 } } { 2 \sigma ^ { 2 } } \right\} \\
    &- \frac { \sqrt { \pi } } { 2 } \frac { ( d - 1 ) [ \mu + ( d - 1 ) ] } { \sigma ^ { 2 } } \exp \left\{ - \frac { \left[ \mu + ( d - 1 ) k_r(0) \right] ^ { 2 } } { 2 \sigma ^ { 2 } } \right\} \Sigma(0)
    \simeq
    \frac { \sigma } { \sqrt { 2 } d k_r(0) } + i \frac { \sigma \Sigma(0) } { \sqrt { 2 } d k_r(0) ^ { 2 } }.
  \end{split}
\end{equation}
We use the differential equation satisfied by the Dawson function $F'(z) + 2zF(z) = 1$ and the asymptotic expansion:
\begin{equation}\label{eq:asymptotic dawson}
  F ( z ) = \sum _ { k = 0 } ^ { \infty } \frac { ( 2 k - 1 )!! } { 2 ^ { k + 1 } z ^ { 2 k + 1 } } = \frac { 1 } { 2 z } + \frac { 1 } { 4 z ^ { 3 } } + \frac { 3 } { 8 z ^ { 5 } } + \cdots\ \mathrm{as}\ |z|\to\infty.
\end{equation}
The left-hand side of Eq.~(\ref{eq:linearized self consistent equation}) is
\begin{equation}
  \begin{split}
    &F \left[ \frac { \mu + ( d - 1 ) k_r(0) } { \sqrt { 2 } \sigma } \right] + F'\left[ \frac { \mu + ( d - 1 ) k_r(0) } { \sqrt { 2 } \sigma } \right] \left[ - i \frac { ( d - 1 ) \Sigma(0) } { \sqrt { 2 } \sigma } \right] \\
    &= F \left[ \frac { \mu + ( d - 1 ) k_r(0) } { \sqrt { 2 } \sigma } \right] + \left\{ 1 - 2 \frac { \mu + ( d - 1 ) k_r(0) } { \sqrt { 2 } \sigma } F \left[ \frac { \mu + ( d - 1 ) k_r(0) } { \sqrt { 2 } \sigma } \right] \right\} \left[ - i \frac { ( d - 1 ) \Sigma(0) } { \sqrt { 2 } \sigma } \right] \\
    &\simeq \frac { 1 } { 2 } \frac { \sqrt { 2 } \sigma } { \mu + ( d - 1 ) k_r(0) } + \frac { 1 } { 4 } \left[ \frac { \sqrt { 2 } \sigma } { \mu + ( d - 1 ) k_r(0) } \right] ^ { 3 } \\
    &- 2 \frac { \mu + ( d - 1 ) k_r(0) } { \sqrt { 2 } \sigma } \left\{ \frac { 1 } { 4 } \left[ \frac { \sqrt { 2 } \sigma } { \mu + ( d - 1 ) k_r(0) } \right] ^ { 3 } + \frac { 3 } { 8 } \left[ \frac { \sqrt { 2 } \sigma } { \mu + ( d - 1 ) k_r(0) } \right] ^ { 5 } \right\}\left[ - i \frac { ( d - 1 ) \Sigma(0) } { \sqrt { 2 } \sigma } \right].
  \end{split}
\end{equation}
The real part of Eq.~(\ref{eq:linearized self consistent equation}) is
\begin{equation}
  \begin{split}
    \frac { 1 } { 2 } \frac { \sqrt { 2 } \sigma } { \mu + ( d - 1 ) k_r(0) } + \frac { 1 } { 4 } \left[ \frac { \sqrt { 2 } \sigma } { \mu + ( d - 1 ) k_r(0) } \right] ^ { 3 } &= \frac { \sigma } { \sqrt { 2 } d k_r(0) } \\
    \frac { 1 } { \mu + ( d - 1 ) k_r(0) } + \frac { \sigma ^ { 2 } } { \left[ \mu + ( d - 1 ) k_r(0) \right] ^ { 3 } } &= \frac { 1 } { d k_r(0) } \\
    \frac { k_r(0) - \mu } { d k_r(0) \left[ \mu + ( d - 1 ) k_r(0) \right] } + \frac { \sigma ^ { 2 } } { \left[ \mu + ( d - 1 ) k_r(0) \right] ^ { 3 } } &= 0.
  \end{split}
\end{equation}
Because $k_r(0) = \mathcal{O}\left(1\right)$ in the large-dimension limit, this is simplified to
\begin{equation}\label{eq:Gaussian real solution}
  \begin{split}
    \frac { k_r(0)  - \mu } { d ^ { 2 } k_r(0) ^ { 2 } } + \frac { \sigma ^ { 2 } } { d ^ { 3 } k_r(0) ^ { 3 } } &= 0 \\
    dk_r(0)^2 - \mu dk_r(0) + \sigma^2 &= 0.
  \end{split}
\end{equation}
Therefore, the real part is 
\begin{equation}
  k_r(0) = \frac { \mu } { 2 } + \frac { 1 } { 2 } \sqrt { \mu ^ { 2 } - 4 \frac { \sigma ^ { 2 } } { d } },
\end{equation}
which is essentially the same as Eq.~(\ref{eq:infinite dimensional solution}).
The imaginary part of Eq.~(\ref{eq:linearized self consistent equation}) is 
\begin{equation}
  \begin{split}
     - 2 \frac { \mu + ( d - 1 ) k_r(0) } { \sqrt { 2 } \sigma } \left[ \frac { 1 } { 4 } \left( \frac { \sqrt { 2 } \sigma } { \mu + ( d - 1 ) k_r(0) } \right) ^ { 3 } + \frac { 3 } { 8 } \left( \frac { \sqrt { 2 } \sigma } { \mu + ( d - 1 ) k_r(0) } \right) ^ { 5 } \right]&\left[ - \frac { ( d - 1 ) \Sigma(0) } { \sqrt { 2 } \sigma } \right]\\
     + \frac { \sqrt { \pi } } { 2 } \exp \left\{ - \frac { \left[ \mu + ( d - 1 ) k_r(0) \right] ^ { 2 } } { 2 \sigma ^ { 2 } } \right\} &= \frac { \sigma \Sigma(0) } { \sqrt { 2 } d k_r(0) ^ { 2 } } \\
    \frac { \sigma } { \sqrt { 2 } } \left\{ \frac { \mu ^ { 2 } + 2 \mu ( d - 1 ) k_r(0) - ( d - 1 ) k_r(0) ^ { 2 } } { d k_r(0) ^ { 2 } \left[ \mu + ( d - 1 ) k_r(0) \right] ^ { 2 } } - \frac { 3 ( d - 1 ) \sigma ^ { 2 } } { \left[ \mu + ( d - 1 ) k_r(0) \right] ^ { 4 } } \right\} \Sigma(0) &= \frac { \sqrt { \pi } } { 2 } \exp \left\{ - \frac { \left[ \mu + ( d - 1 ) k_r(0) \right] ^ { 2 } } { 2 \sigma ^ { 2 } } \right\}.
  \end{split}
\end{equation}
Employing the same procedure as in the real part, this can be simplified to 
\begin{equation}\label{eq:simplified gaussian imaginary}
  \begin{split}
    \frac { \sigma } { \sqrt { 2 } } \left[ \frac { 2 \mu d k_r(0) - d k_r(0) ^ { 2 } } { d ^ { 3 } k_r(0) ^ { 4 } } - \frac { 3 d \sigma ^ { 2 } } { d ^ { 4 } k_r(0) ^ { 4 } } \right]\Sigma(0) &= \frac { \sqrt { \pi } } { 2 } \exp \left[ - \frac { d ^ { 2 } k_r(0) ^ { 2 } } { 2 \sigma ^ { 2 } } \right] \\
    \left[ 2 \mu d k_r(0) - d k_r(0) ^ { 2 } - 3 \sigma ^ { 2 } \right] \Sigma(0) &= \frac{d ^ { 3 } k_r(0) ^ { 4 }}{\sigma} \sqrt { \frac { \pi } { 2  } } \exp \left[ - \frac { d ^ { 2 } k_r(0) ^ { 2 } } { 2 \sigma ^ { 2 } } \right]\\
    \left[k_r(0) - 2 \frac { \sigma ^ { 2 } } { \mu d } \right] \Sigma(0) &= \frac { d ^ { 2 } k_r(0) ^ { 4 } } { \mu \sigma } \sqrt { \frac { \pi } { 2 } } \exp \left[ - \frac { d ^ { 2 } k_r(0) ^ { 2 } } { 2 \sigma ^ { 2 } } \right] \\
    \Sigma(0) &= \sqrt { \frac { \pi } { 2 } } \frac { d ^ { 2 } k_r(0) ^ { 4 } \exp \left[ - \frac { d ^ { 2 } k_r(0) ^ { 2 } } { 2 \sigma ^ { 2 } } \right]} { \mu \sigma \left[k_r(0) - 2 \frac { \sigma ^ { 2 } } { \mu d } \right]}. 
  \end{split}
\end{equation}
To obtain the third line from the second, we have used Eq.~(\ref{eq:Gaussian real solution}).

\section{Marginal solution by local instability}\label{sec:Marginal solution by the local instability}

In this appendix, we solve Eq.~(\ref{eq:equation for the real part}) in the lowest-frequency region.
The left-hand side of Eq.~(\ref{eq:equation for the real part}) can be transformed as follows:
\begin{equation}\label{eq:principal value}
  \begin{split}
    &\mathcal{P}\int_{-\Delta}^{\Delta}\frac{dk_\alpha P(k_{\alpha}+\mu)}{k_\alpha + \mu + (d-1)k_r(\omega) - dA_d\omega^2} \\
    &=\mathcal{P}\int_{-1}^{1}\frac{dk_\alpha \Delta P(\Delta k_{\alpha}+\mu)}{\Delta k_\alpha + \mu + (d-1)k_r(\omega) - dA_d\omega^2} \\
    &\equiv\mathcal{P}\int_{-1}^1\frac{dx\tilde{P}(x)}{\Delta k_\alpha + \mu + (d-1)k_r(\omega) - dA_d\omega^2}. \\
  \end{split}
\end{equation}
$\tilde{P}(x)$ is the same distribution as $P(x)$, but it is supported on the interval $[-1,1]$.
$k_r(\omega)$ is decomposed to $k_r(\omega)=\left[-\mu+\Delta + \delta k_r(\omega)\right]/(d-1)$, and the denominator of the integrand becomes
\begin{equation}
  \frac{\Delta + \delta k_r(\omega) -dA_d\omega^2}{\Delta} = 1 - \frac{-\delta k_r(\omega) + dA_d\omega^2}{\Delta} \equiv 1 - \frac{\eta(\omega)}{\Delta}.
\end{equation}
We assume that $\delta k_r(\omega), \eta(\omega) \sim \omega^2$ as $\omega\to0$.
In the following, Eq.~(\ref{eq:principal value}) is expanded to the first order in $\omega^2$.
Before that, we consider the principal value part of Eq.~(\ref{eq:principal value}).
Equation~(\ref{eq:principal value}) can be expressed as follows:
\begin{equation}\label{eq:decomposed principal value}
  \begin{split}
    &\frac{1}{\Delta}\mathcal{P}\int_{-1}^{1}\frac{dx \tilde{P}(x)}{x + 1-\eta(\omega)/\Delta} \\
    &= \frac{1}{\Delta}\mathcal{P}\int_{-1}^{-1+2\eta(\omega)/\Delta}\frac{dx \tilde{P}(x)}{x + 1-\eta(\omega)/\Delta} + \frac{1}{\Delta}\int_{-1+2\eta(\omega)/\Delta}^{1}\frac{dx \tilde{P}(x)}{x + 1-\eta(\omega)/\Delta}. \\
  \end{split}
\end{equation}
Using the assumption $\tilde{P}(x) = (\mathrm{const.})(x+1)^{\nu}$ with $\nu>1$ at $x\simeq-1$, the first term can be proved to be negligible as follows:
\begin{equation}
  \begin{split}
    &\frac{1}{\Delta}\mathcal{P}\int_{-1}^{-1+2\eta(\omega)/\Delta}\frac{dx \tilde{P}(x)}{x + 1-\eta(\omega)/\Delta} \\
    &= \frac{\eta(\omega)^\nu}{\Delta^{\nu+1}}\mathcal{P}\int_{-1}^{-1+2\eta(\omega)/\Delta}\frac{dx\Delta/\eta(\omega) \left[(x+1)\Delta/\eta(\omega)\right]^\nu}{(x + 1)\Delta/\eta(\omega)-1} \\
    &= \frac{\eta(\omega)^\nu}{\Delta^{\nu+1}}\mathcal{P}\int_{0}^{2}dt\frac{ t^\nu}{t-1}, \\
    &= {o}(\omega^2),
  \end{split}
\end{equation}
where the numerical factors are ignored during calculation.
Now, we expand Eq.~(\ref{eq:principal value}) up to the first order in $\omega^2$ as follows:
\begin{equation}\label{eq:left hand side}
  \begin{split}
    &\frac{1}{\Delta}\mathcal{P}\int_{-1}^{1}\frac{dx \tilde{P}(x)}{x + 1-\eta(\omega)/\Delta}\\
    &= \frac{1}{\Delta}\int_{-1+2\eta(\omega)/\Delta}^{1}\frac{dx \tilde{P}(x)}{x + 1 - \eta(\omega)/\Delta} + {o}(\omega^2)\\   
    &= \frac{1}{\Delta}\int_{-1}^{1}\frac{dx \tilde{P}(x)}{x + 1} + \frac{1}{\Delta^2}\int_{-1}^1dx\frac{\tilde{P}(x)}{(x+1)^2}\left[-\delta k_r(\omega) + dA_d\omega^2\right] + {o}(\omega^2)\\    
    &\equiv \frac{I_1}{\Delta} + \frac{I_2}{\Delta^2}\left[-\delta k_r(\omega) + dA_d\omega^2\right] + {o}(\omega^2),\\
  \end{split}
\end{equation}
where 
\begin{equation}
  I_{m} \equiv \int_{-1}^1dx\frac{\tilde{P}(x)}{(x+1)^m}.
\end{equation}
The right-hand side of Eq.~(\ref{eq:equation for the real part}) can also be expanded as 
\begin{equation}\label{eq:right hand side}
  \begin{split}
    &\frac{1}{dk_r(\omega)}\left[1 + \frac{A_d\omega^2}{k_r(\omega)}\right] \\    
    &= \frac{d-1}{d\left[-\mu + \Delta + \delta k_r(\omega)\right]}\left[ 1 + \frac{(d-1)A_d\omega^2}{-\mu + \Delta + \delta k_r(\omega)}\right] \\ 
    &= \frac{d-1}{d(-\mu+\Delta)} + \frac{(d-1)}{d(-\mu+\Delta)^2}\left[ - \delta k_r(\omega) + (d-1)A_d\omega^2\right] + \mathcal{O}(\omega^4). \\
  \end{split}
\end{equation}
Comparing Eqs.~(\ref{eq:left hand side}) and~(\ref{eq:right hand side}), the equation at zero frequency is 
\begin{equation}\label{eq:equation at zero frequency}
  I_1 = \frac{\Delta(d-1)}{d(-\mu+\Delta)}.
\end{equation}
The equation for $\delta k_r(\omega)$ is 
\begin{equation}
  \begin{split}
    \frac{I_2}{\Delta^2}\left[-\delta k_r(\omega) + dA_d\omega^2\right] &= \frac{(d-1)}{d(-\mu+\Delta)^2}\left[ - \delta k_r(\omega) + (d-1)A_d\omega^2\right] \\
    I_2\left[-\delta k_r(\omega) + dA_d\omega^2\right] &= \frac{d}{d-1}I_1^2\left[ - \delta k_r(\omega) + (d-1)A_d\omega^2\right] \\
    \frac{\delta k_r(\omega)}{d-1} &= -\frac{dI_2 - dI_1^2}{dI_1^2 - (d-1)I_2}A_d\omega^2. \\
  \end{split}
\end{equation}
To obtain the second line from the first, we have used Eq.~(\ref{eq:equation at zero frequency}).

\section{Model with continuously vanishing stiffness}\label{sec:Model with the continuously vanishing stiffness}

\begin{figure}
    \centering
    \includegraphics{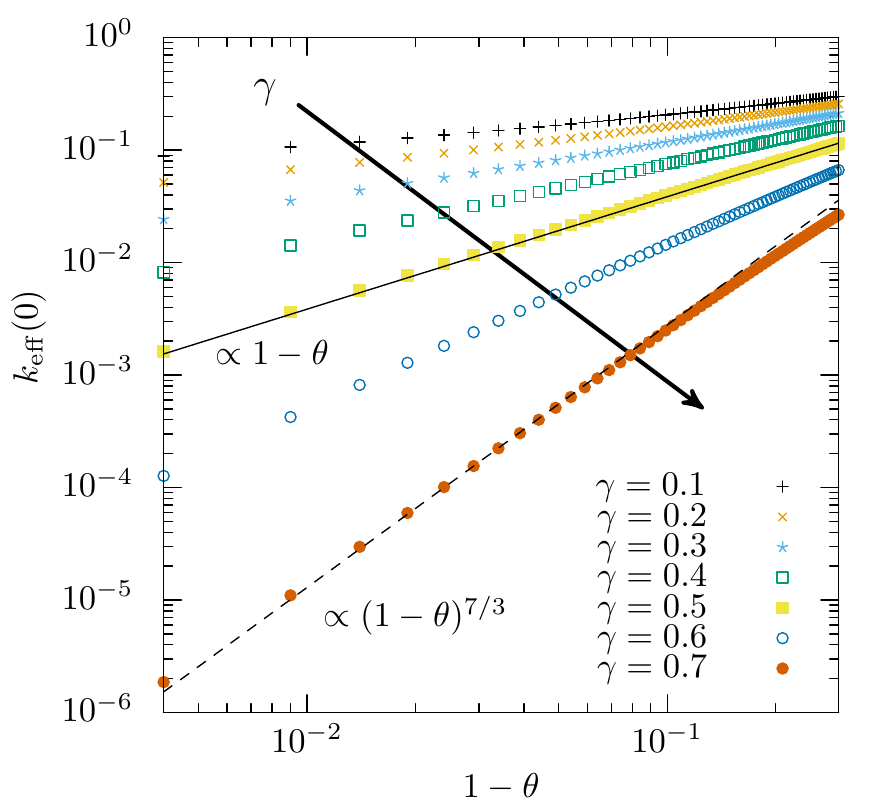}
    \caption{
    Effective stiffness at zero frequency for the model with Eq.~(\ref{eq:distribution with gamma}).
    The Solid line is proportional to $1-\theta$ and the dotted line is proportional to $(1-\theta)^{7/3}$, both of which are predicted in Eq.~(\ref{eq:solution for gamma}).
    }
    \label{fig:fig5}
\end{figure}

In the analysis of the SDM, we implicitly assume that the effective stiffness at zero frequency does not vanish, $k_{\mathrm{eff}}(0)\neq0$.
However, when the probability distribution $P(k_\alpha)$ has a divergence at $k_\alpha=0$, the effective stiffness can be zero.
The divergent distribution can be observed in jammed systems, with their initial stress ignored~\cite{mizuno2016elastic}.
Because the QLVs disappear when we ignore the initial stress~\cite{Mizuno2017Continuum,Lerner2018Frustration}, it is noteworthy to consider the divergent distributions.
The most notable example is the percolation problem~\cite{Feng1984Percolation,Feng1985Effective,Wyart2010Scaling,During2013Phonon} with the Bernoulli distribution 
\begin{equation}
    P(k_\alpha) = p\delta(k_\alpha-1) + (1-p)\delta(k_\alpha).
\end{equation}
In this case, the effective stiffness vanishes linearly as follows:
$k_{\mathrm{eff}}(0)\sim p-\theta$.

Here, we present another probability distribution that yields a vanishing stiffness.
We analyze
\begin{equation}\label{eq:distribution with gamma}
    P(k_\alpha) =
    \begin{cases}
        Ck_\alpha^{-\gamma} & k_\alpha\in\left[0,k_{\mathrm{max}}\right] \\
        0 & \mathrm{otherwise}
    \end{cases}
    ,
\end{equation}
where $0<\gamma<1$ and $C = (1-\gamma)k_{\mathrm{max}}^{\gamma-1}$.
Considering the zero-frequency limit, Eq.~(\ref{eq:self consistent equation for the VDM}) becomes
\begin{equation}\label{eq:self consistent equation for gamma}
    \overline{\frac{1}{k_\alpha + (1-\theta)k_{\mathrm{eff}}(0)/\theta}} = \frac{\theta}{k_{\mathrm{eff}}(0)}.
\end{equation}
In this model, $\theta$ is considered as a control parameter, and we can show that $k_{\mathrm{eff}}(0)\to0$ as $\theta\to1$.
The left-hand side of the equation is approximated as
\begin{equation}
    \begin{split}
        &\overline{\frac{1}{k_\alpha + (1-\theta)k_{\mathrm{eff}}(0)/\theta}} \\
        &= C\int_0^{k_{\mathrm{max}}}\frac{k_\alpha^{-\gamma}dk_\alpha}{k_\alpha + (1-\theta)k_{\mathrm{eff}}(0)/\theta} \\
        &= C\left[\frac{1-\theta}{\theta}k_{\mathrm{eff}}(0)\right]^{1-\gamma}\int_0^{k_{\mathrm{max}}\theta/(1-\theta)k_{\mathrm{eff}}(0)}\frac{t^{-\gamma}}{t+1} dt \\
        &\simeq C\left[\frac{1-\theta}{\theta}k_{\mathrm{eff}}(0)\right]^{1-\gamma}\int_0^\infty \frac{t^{-\gamma}}{t+1} dt.
    \end{split}
\end{equation}
Therefore, the effective stiffness at zero frequency is
\begin{equation}\label{eq:solution for gamma}
    k_{\mathrm{eff}}(0) \sim (1-\theta)^{\frac{\gamma}{1-\gamma}}.
\end{equation}
To validate this analytical prediction, we numerically solved Eq.~(\ref{eq:self consistent equation for gamma}).
Figure~\ref{fig:fig5} presents the effective stiffness as a function of $1-\theta$ with $\gamma=0.1,\ 0.2,\ 0.3,\ 0.4,\ 0.5,\ 0.6$, and $0.7$.
The solid line indicates Eq.~(\ref{eq:solution for gamma}) with $\gamma=0.5$, and the dotted line indicates Eq.~(\ref{eq:solution for gamma}) with $\gamma=0.7$.

\twocolumngrid

\bibliography{main}

\end{document}